\documentclass[preprint,12pt,authoryear]{elsarticle}

\usepackage{amssymb}
\usepackage{caption}
\usepackage{subcaption}
\usepackage{booktabs}
\usepackage{float}
\usepackage{comment}	
\usepackage{lineno}

\journal{Icarus}

\begin{document}

\begin{frontmatter}



\title{Analysis of CN emission as a marker of organic compounds in meteoroids using laboratory simulated meteors}


\author[inst1]{Adriana Pisarčíková}\ead{pisarcikova@fmph.uniba.sk}

\affiliation[inst1]{organization={Faculty of Mathematics, Physics and Informatics, Comenius University in Bratislava},
            addressline={Mlynská dolina}, 
            city={Bratislava},
            postcode={84248}, 
            country={Slovakia}}

\author[inst1]{Pavol Matlovič}
\author[inst1]{Juraj Tóth}
\author[inst2]{Stefan Loehle}
\author[inst3]{Ludovic Ferri\`ere}
\author[inst2]{David Leiser}
\author[inst2]{Felix Grigat}
\author[inst4]{Jérémie Vaubaillon}

\affiliation[inst2]{organization={High Enthalpy Flow Diagnostics Group, Institute of Space Systems, University of Stuttgart},
            addressline={Pfaffenwaldring 29}, 
            city={Stuttgart},
            postcode={70569}, 
            country={Germany}}
            
\affiliation[inst3]{organization={Natural History Museum Vienna},
            addressline={Burgring 7}, 
            city={Vienna},
            postcode={1010}, 
            country={Austria}}
                        
\affiliation[inst4]{organization={IMCCE, Observatoire de Paris},
            addressline={PSL, 77 Av Denfert Rochereau}, 
            city={Paris},
            postcode={75014}, 
            country={France}}

\begin{abstract}
Fragments of small solar system bodies entering Earth's atmosphere have possibly been important contributors of organic compounds to the early Earth. The cyano radical (CN) emission from meteors is considered as potentially one of the most suitable markers of organic compounds in meteoroids, however, its detection in meteor spectra has been thus far unsuccessful. With the aim to improve our abilities to identify CN emission in meteor observations and use its spectral features to characterize the composition of incoming asteroidal meteoroids, we present a detailed analysis of CN emission from high-resolution spectra of 22 laboratory simulated meteors including ordinary, carbonaceous, and enstatite chondrites, as well as a large diversity of achondrites (i.e., ureilite, aubrite, lunar, martian, howardite, eucrite, and diogenite), mesosiderite, and iron meteorites. We describe the variations of CN emission from different classes of asteroidal meteor analogues, its correlation and time evolution relative to other major meteoroid components. We demonstrate that CN can be used as a diagnostic spectral feature of carbonaceous and carbon-rich meteoroids, while most ordinary chondrites show no signs of CN. Our results point out strong correlation between CN and H emission and suggest both volatile features are suitable to trace contents of organic matter and water molecules present within meteoroids. For the application in lower resolution meteor observations, we demonstrate that CN can be best recognized in the early stages of ablation and for carbon-rich materials by measuring relative intensity ratio of CN band peak to the nearby Fe I-4 lines.
\end{abstract}


\begin{highlights}
\item First analysis of CN emission from various ablated meteorites
\item CN emission identified as a diagnostic spectral feature of carbon-rich meteorites
\item CN and H emission linked to organic matter and water content
\item Method for identification of CN in lower resolution meteor spectra proposed
\end{highlights}

\begin{keyword}
astrobiology \sep spectroscopy \sep meteorite
\PACS 0000 \sep 1111
\MSC 0000 \sep 1111
\end{keyword}

\end{frontmatter}

\section{Introduction}
\label{sec:intro}

The various abundances of organic matter found in asteroids and comets originate from the formation processes of the interstellar medium \citep{1993A&A...273..583J, 1998ApJ...498L..83L, 2004come.book..105L}. Small solar system bodies are assumed to have been responsible for the emergence of prebiotic molecules necessary for the origin of life on Earth \citep{Oro, 1992Natur.355..125C, 1997coel.conf..147C, 1992OLEB...21..279D}. The impacting interplanetary material is considered to be one of the main contributors of organic molecules to early Earth \citep{2000lsr..book...57J, 2001ESASP.495..247J, article}.

While organic matter is abundantly present in all comets and spectral studies can focus on revealing the variations in the contents of different compounds, the abundance of organic matter in different types of asteroids remains an open question \citep{2020SSRv..216...54M}. Studies of the fragments of asteroids and comets -- meteoroids -- which continuously enter the Earth's atmosphere from various sources in the solar system can provide important spectral data to help tackle this issue. Meteoroids ablate in the Earth's atmosphere as meteors emitting strong radiation. Analyzing the emission spectra allows to gain detailed information about the atoms and molecules present in the meteoroid and interacting with the surrounding air \citep{1993A&A...279..627B, 2016Icar..278..248B}.

A suitable trace feature for the detection of organic compounds in small solar system bodies captured through meteor observations appears to be the cyano radical (CN). CN has been detected by remote optical observations in several comets since the 19th century \citep{10.2307/108925, 1881Obs.....4..252D}. In recent years, even the first in-situ detection of CN in the coma of comet 67P/Churyumov–Gerasimenko has taken place \citep{2020MNRAS.498.2239H}.

The origin of the CN observed in the cometary coma has been long associated with hydrogen cyanide (HCN) as the sole source. However, in the last few decades, two main CN sources have been considered: CN-bearing refractories (HCN polymers, hexamethylenetetramine (HMT), tholin, CHON dust grains) and CN-bearing volatiles \citep{2020MNRAS.498.2239H} as the dominant source of CN production (mainly HCN, cyanogen (C\textsubscript{2}N\textsubscript{2}), cyanoacetylene (HC\textsubscript{3}N), acetonitrile (CH\textsubscript{3}CN)). Refractories carrying CN are assumed to be released from the interior of the comet along with dust particles and could generate HCN or CN radicals. CN products derived from volatile CN-bearing species are formed during the photodissociation of these species.

There has been several efforts to detect CN in meteor spectra in the past decades (see e.g., \citet{1971BAICz..22..219C, 1998EM&P...82...71R, article, 2016Icar..278..248B}). The detection was unsuccessful likely due to the typically insufficient resolution of meteor spectrographs, which are unable to resolve the CN band from the strong emission of surrounding Fe I lines. Because of its strong B → X transition of low excitation energy \citep{1998EM&P...82...71R}, the CN emission is the most suitable tracer of organic compounds in the visible and near-UV range meteor spectra. At typical meteoric temperatures and instrumental resolutions, this vibrational band structure peaks at around 388.3 nm \citep{1998EM&P...82...71R}. CN emission from meteor spectra is expected to originate either directly from the meteoroid composition and it may be generated from reactions with N bound in organic matter, or due to the interaction of the meteoric C atoms with molecular N\textsubscript{2} originating in the atmosphere \citep{article}. 

In this work we present an analysis of the CN emission in spectra of different types of meteorites tested in a plasma wind tunnel simulating meteoric conditions. The fitting of the CN band was previously done in the terrestrial rock argillite tested under the same laboratory conditions \citep{2017ApJ...837..112L}. We provide the first overview of the presence, relative intensity, and time evolution of the CN emission in different meteorites representing a wide range of asteroidal materials. This way, we aim to indicate CN as a suitable tracer of organic matter in meteoroids, demonstrate the detectable variations of organic matter in different asteroidal materials, and help constrain the instrumental limits for an efficient detection of CN in meteors.

First, in Section \ref{experiments}, we describe the laboratory conditions and instrumentation used for the meteorite ablation tests in the plasma wind tunnel and the data processing methodology. The following Section \ref{results} contains our results of a detailed study of the CN emission in spectra of different meteorites. In this section we focus on an analysis of the presence of CN and H$\alpha$ in meteorite spectra and their mutual correlation, and an analysis of the relative intensity and time evolution of the CN band emission based on monochromatic light curves. The conclusions derived from the obtained results are summarized in Section \ref{conclusions}.

\section{Laboratory experiments and methods}
\label{experiments}
 \citet{2017ApJ...837..112L} established an experimental setup in an arc-jet wind tunnel facility suitable for the analysis of meteoroid entry physics. Overall, three experimental campaigns (2020-2022) were performed within a cooperation between the Comenius University in Bratislava, Slovakia (CUB) and the High Enthalpy Flow Diagnostics Group at the Institute of Space Systems, University of Stuttgart, Germany (HEFDiG). The following analysis is based on the measurement of spectra of 22 meteorite samples simultaneously captured by the high-resolution HEFDiG Echelle spectrograph and the spectrograph AMOS-Spec-HR from the CUB, which is used within the global AMOS (All-sky Meteor Orbit System) network \citep{2015P&SS..118..102T} for observing spectra of meteors in the Earth's atmosphere \citep{2019A&A...629A..71M, 2020A&A...636A.122M}. Given the relatively low resolution of AMOS-Spec-HR, we only use these data to evaluate the possibility to recognize CN in corresponding low resolution spectra. The analysis of relative intensities and time evolution of CN emission from different meteorite types is based on the detailed Echelle spectra. The uniquely large dataset of tested meteorites allows us to examine the presence of CN in almost all major meteorite classes including different ordinary chondrites, enstatite chondrites, carbonaceous chondrites, achondrites and mesosiderite group of stony irons. These types represent the most abundant meteorite falls.
 
\subsection{Experiment conditions and instrumentation}
The Institute of Space Systems of the University of Stuttgart operates several plasma wind tunnels \citep{loehle2021a}, which were developed in the early 1980s for basic testing of thermal protection materials required for spacecraft to safely enter the atmosphere of planets. The meteorite experiments were carried out in the Plasma Wind Tunnel 1 (PWK1) with a plasma flow condition with local mass-specific enthalpy of 70 MJ kg\textsuperscript{-1} at a stagnation pressure of $\sim$24\,hPa. This corresponds to the entry of a meteoroid with a diameter of $\sim$4\,cm at an altitude of $\sim$80\,km in the Earth's atmosphere, with an assumed meteoroid entry velocity of $\sim$12\,km s\textsuperscript{-1}. This plasma flow condition was used for the first tests with meteorite samples \citep{2017ApJ...837..112L, 2018A&A...613A..54D}.

The Echelle spectrograph of HEFDiG is a fiber-fed system providing a wavelength range of 250–880 nm \citep{2017ApJ...837..112L}. From the lower to the upper end of this spectral interval, the spectral dispersion varies from 43 pm px\textsuperscript{-1} to 143 pm px\textsuperscript{-1} (resolving power R $\approx$ 10 000). An Echelle high-order diffraction grating of 300 grooves per millimeter (gpmm) is utilized at orders 40–60. Another mounted diffraction grating with higher diffraction of about 1000 gpmm causes dispersion and alignment of obtained spectra. As a result, a high resolution and long wavelength interval is obtained.

The AMOS-Spec-HR system provides an image resolution of 2048 x 1536 px (1.76 arcmin px\textsuperscript{-1}) and a resulting field of view (FOV) of 60° x 45° and a frame rate of 15 fps. The essential components of this spectrograph are 6 mm f/1.4 lens and a digital camera. The setup of holographic diffraction grating with 1000 gpmm provides a dispersion of 0.5 nm px\textsuperscript{-1} (resolving power R $\approx$ 550). The spectral system allows to analyze spectral events in the visual spectrum range of approximately 370–900 nm.

\begin{figure}
\centerline{\includegraphics[width=.90\columnwidth,angle=0]{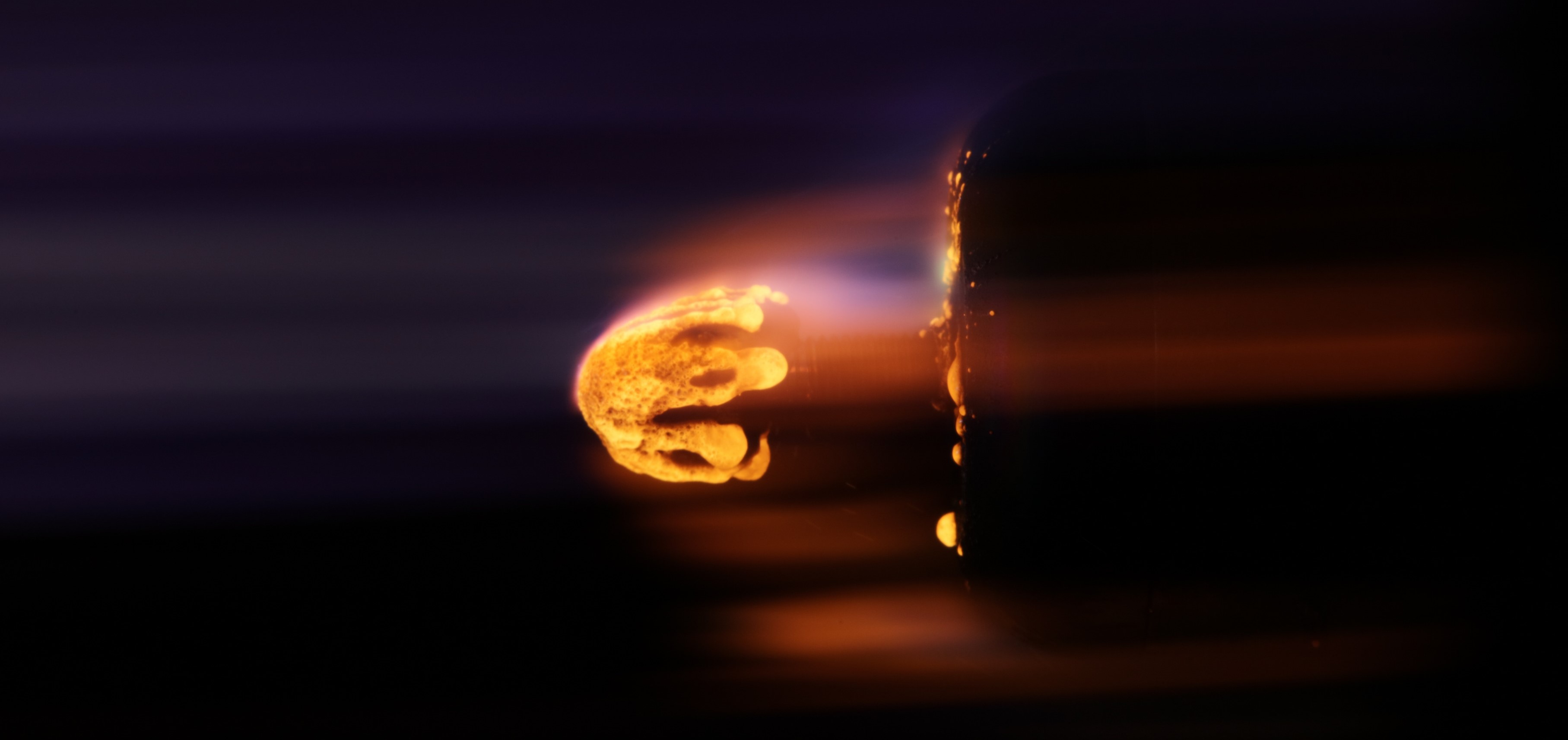}}
\caption{Ablation of Chelyabinsk meteorite sample in PWK1 plasma wind tunnel at the HEFDiG of the Institute of Space Systems at the University of Stuttgart.} 
\label{ablation}
\end{figure}

The large selection of tested meteorite samples, obtained from and in collaboration with the Natural History Museum Vienna, mainly consist of meteorite falls rather than finds to limit terrestrial contamination. Meteorite samples were cut into 1 cm diameter cylinders with lengths varying from $\sim$1\ to 2 cm or into $\sim$1\ cm diameter cubes depending on the availability and fragility of the samples. These dimensions are required for accurate experiment conditions with respect to the mentioned entry conditions. The meteorite samples were attached to a copper stick mounted on a standard ESA (European Space Agency) probe holder on a four-axis moving platform inside a 6 m long and 2 m wide vacuum chamber of the PWK1 plasma wind tunnel during experiment. The PWK1 plasma wind tunnel was evacuated, and subsequently, the magnetoplasmadynamic generator was turned on. The moving platform was used to transport the probe held outside the plasma flow to its direction after the air flow is stabilized. The duration of exposure of the sample to the air plasma flow ranged around 3-12 s depending on the meteorite composition and durability of the meteorite holder. For some smaller samples or samples with higher risk of fragmentation upon drilling for the copper stick holder, a high-temperature ceramic glue Resbond 940HT was used to attach the sample to the copper holder. The spectrum of a pure glue sample was obtained to ensure that the contamination from the glue to the obtained spectra was negligible, as was confirmed. An example of a melting Chelyabinsk meteorite sample in the PWK1 plasma wind tunnel is shown in Fig. \ref{ablation}.

\subsection{Data processing}
The calibration of the Echelle spectra of the ablated meteorites was performed after each laboratory experiment using a calibration lamp located at the position where the meteorite sample was previously placed. The radiation of the calibration lamp measured in the laboratory environment is used to convert the ADU camera units (Analog-to-Digital Units) to spectral radiance. Considering the typical case of the meteorite ablation of $\sim$4\,s duration and the effort to obtain the highest possible camera gain, 15-70 frames of the meteorite emission spectrum are recorded. The resulting emission spectrum was obtained by summing the intensity profiles of the individual calibrated frames. The last step before calculating line intensities is subtracting spectral baselines using the Fityk program \citep{Wojdyr:ko5121}. Within this program, a synthetic spectrum consisting of the main emission multiplets of the meteor was modeled and then fitted to the calibrated spectrum using the damped least-squares method (the Levenberg--Marquardt algorithm) to measure the relative intensities of spectral emission lines. For the shape of all modeled lines in the synthetic spectrum, Gaussian line profiles were used with appropriate full width at half maximum (FWHM) adjusted by automatic fit, typically $\sim$0.1\,nm. The error bars calculation of the measured line intensities was estimated based on the signal to noise ratio (SNR) in each meteor spectrum. The multiplet numbers used in this work are taken from \citet{1945CoPri..20....1M}.

The spectral data analysis of AMOS data was carried out according to the procedure described in \citet{10.1093/mnras/stac927}. Each meteorite spectrum was corrected for noise, other sources of illumination and spectral sensitivity of the system and later manually scanned in individual frames of the video recording. The resulting meteorite spectrum was obtained by summing all intensity profiles and scaled using well-known lines and a polynomial fit of the third order.

\begin{figure}
\centerline{\includegraphics[width=.95\columnwidth,angle=0]{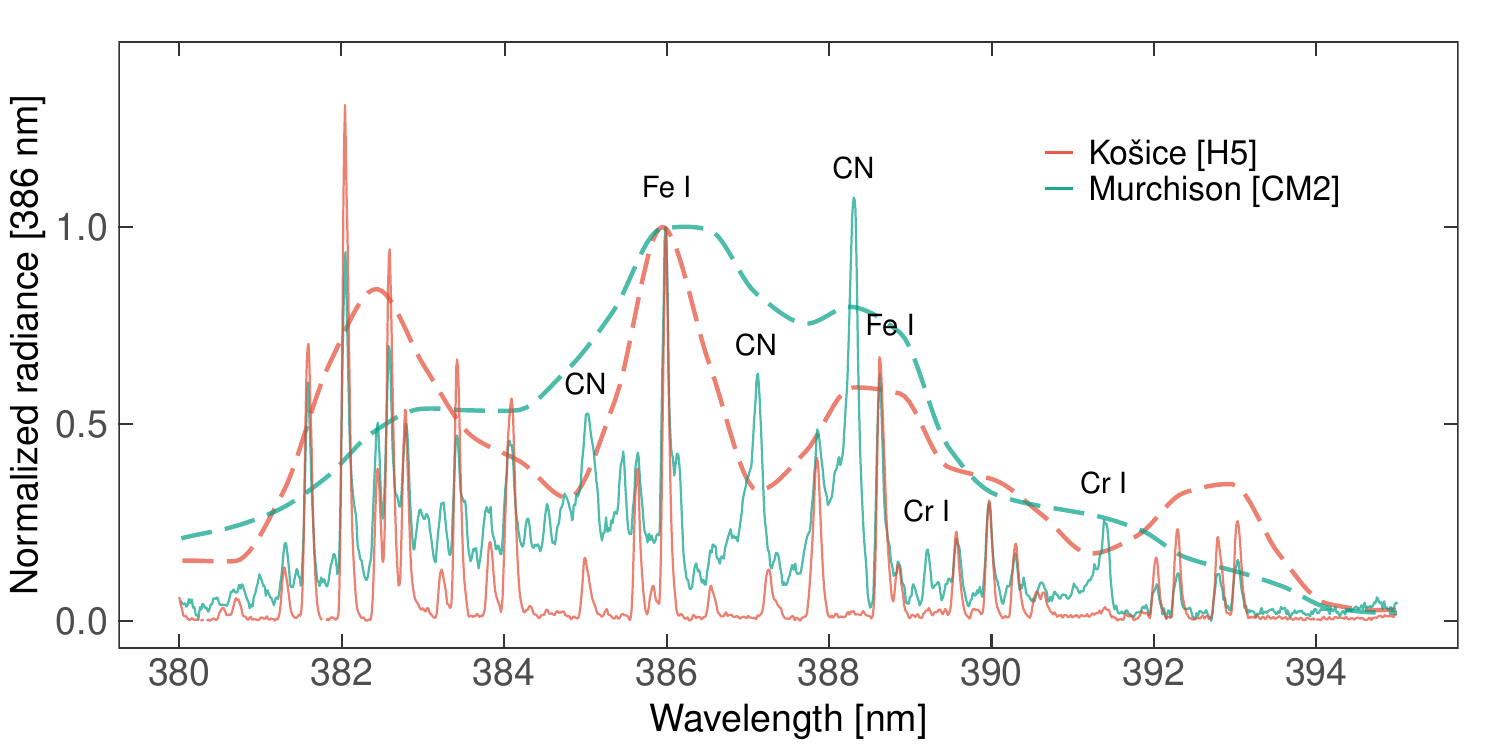}}
\caption{Calibrated emission spectra of two meteorite samples in the 380-395 nm region demonstrating the difference in resolution of two instruments: Echelle spectrograph with R $\approx$ 10 000 (solid line) and AMOS-Spec-HR with R $\approx$ 550 (dashed line). Spectra of the ordinary chondrite (H5) Košice with the absence of the CN band and carbonaceous chondrite (CM2) Murchison with strong CN emission are displayed. The spectra are normalized to unity at the peak intensity of the 386.0 nm Fe I line.} 
\label{profiles}
\end{figure}

In this work, we analyzed the presence and relative intensity of the CN band in emission spectra of tested meteorites. The strongest peak of this band is located near 388.3 nm (studied in our analysis), followed by weaker CN peaks near 387.1 nm and 385.0 nm. Due to the high-resolution (R $\approx$ 10 000) of the Echelle data, the measurement of CN intensity in the spectra of ablated meteorites is straightforward, while in the case of the lower resolution (R $\approx$ 550) of the AMOS data, intensity of the CN band is affected by contributions of surrounding Fe lines. The comparison of the emission spectrum of Murchison meteorite in B → X CN band region in lower resolution data of AMOS and higher resolution Echelle data is displayed in Fig. \ref{profiles}.

\begin{figure}
\centerline{\includegraphics[width=.95\columnwidth,angle=0]{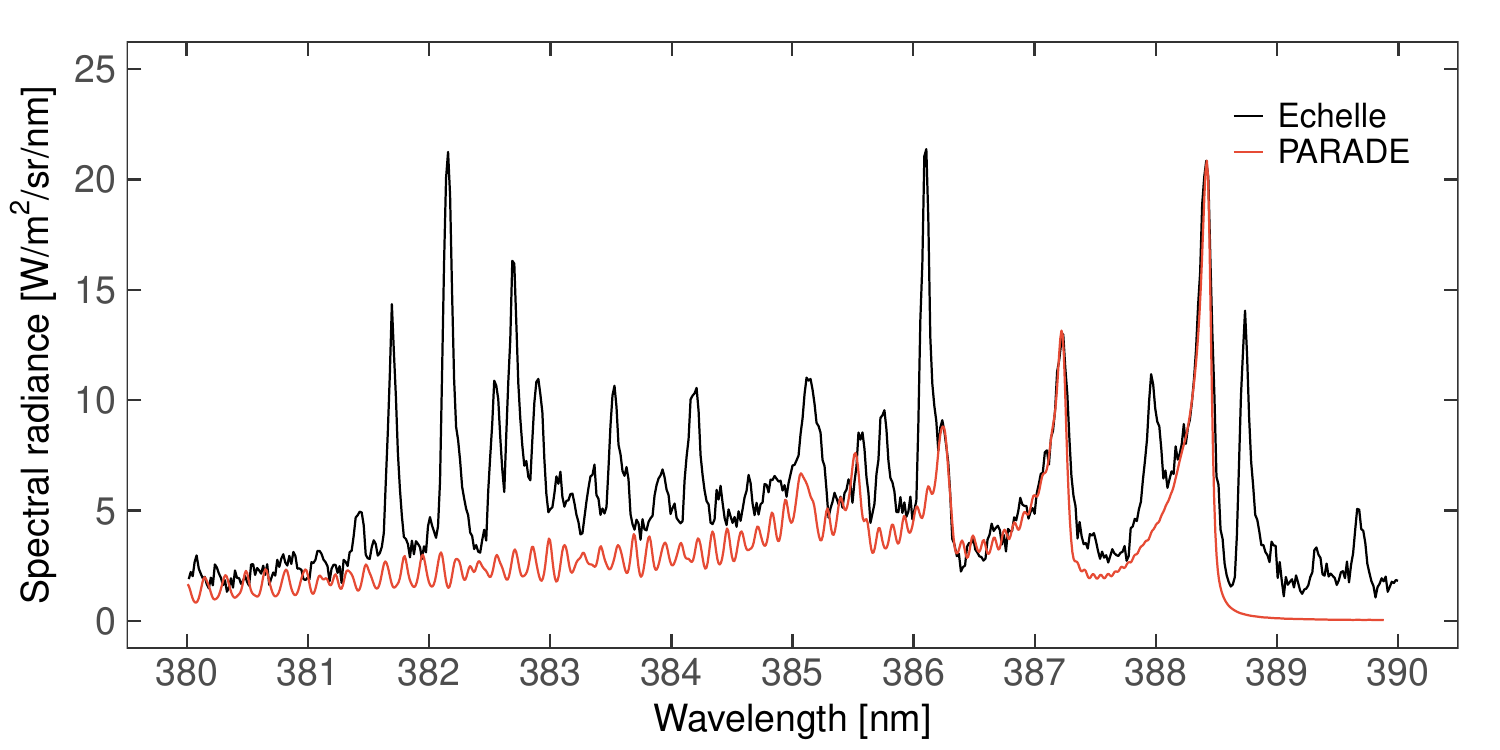}}
\caption{The model of the CN emission band fitted to the spectrum of the carbonaceous chondrite (CM2) Murchison. An average spectrum from five frames captured by the Echelle spectrograph was used. The CN model was obtained using the line-by-line emission code PARADE at rotational and vibrational temperatures T\textsubscript{rot} = 6500 K and T\textsubscript{vib} = 6500 K, respectively.} 
\label{MUR_CN_fit}
\end{figure}

An illustration of the model of the CN (B → X) $\Delta\nu$ = 0 band fitted to an observed spectrum of the Murchison meteorite is displayed in Fig. \ref{MUR_CN_fit}. The CN model was obtained using the line-by-line emission code PARADE \citep{parade2009, 2021M&PS...56..352L}. Equilibrium was assumed between the translational, rotational,  vibrational and electronic temperatures, which were manually varied to fit the simulated spectrum to the data recorded with the Echelle spectrometer. In order to reduce the effect of noise on the fit, the spectra of five successive frames were averaged for each temperature estimate. The fit of the CN band displayed in Fig. \ref{MUR_CN_fit} was obtained at the resulting rotational T\textsubscript{rot} = 6500 K and vibrational temperatures T\textsubscript{vib} = 6500 K.

\section{Results}
\label{results}
\subsection{The detection of CN in meteorite spectra and correlation with H} \label{H and CN emission}

\begin{figure}
\centerline{\includegraphics[width=0.8\columnwidth,angle=0]{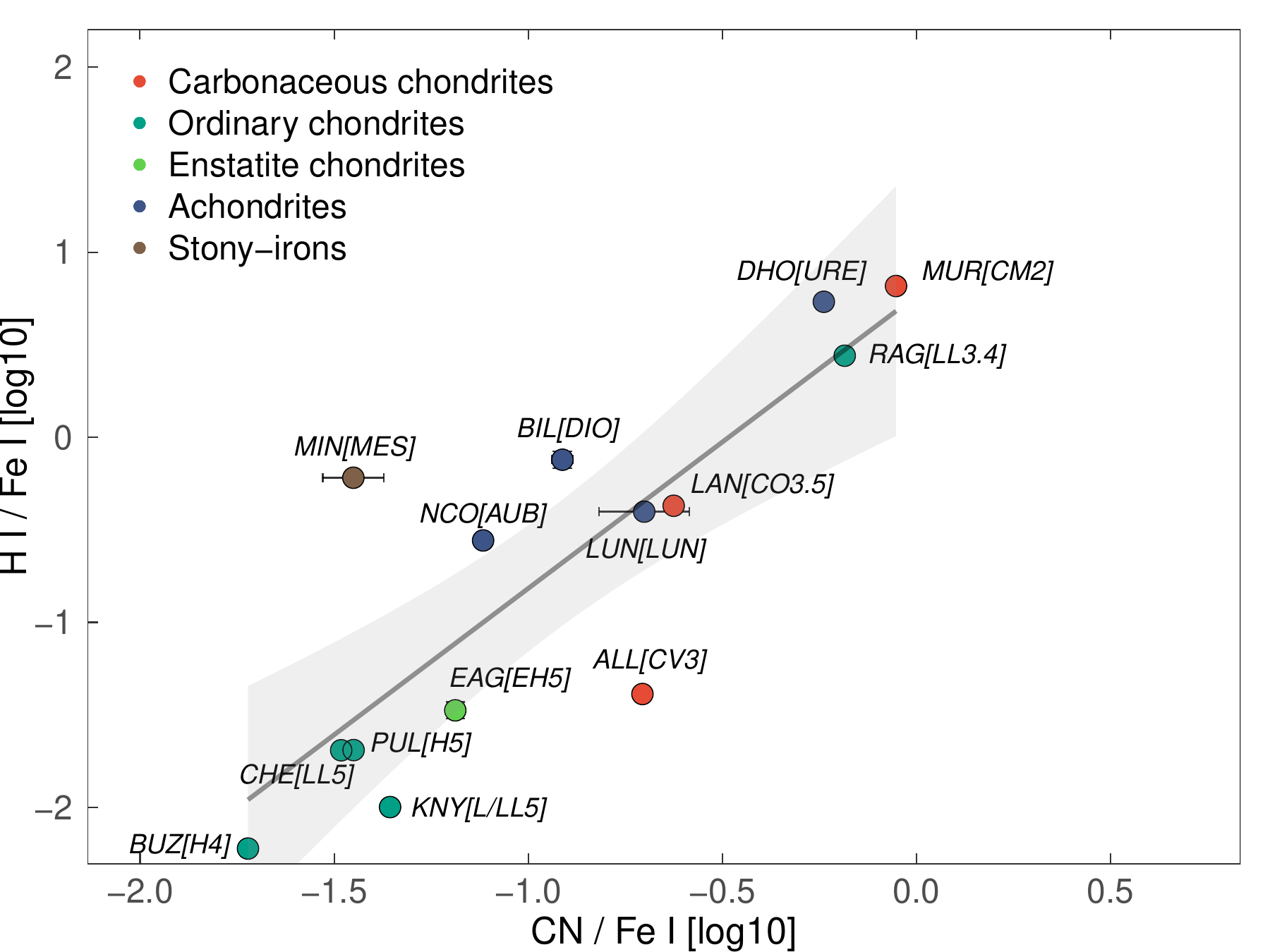}}
\centerline{\includegraphics[width=0.8\columnwidth,angle=0]{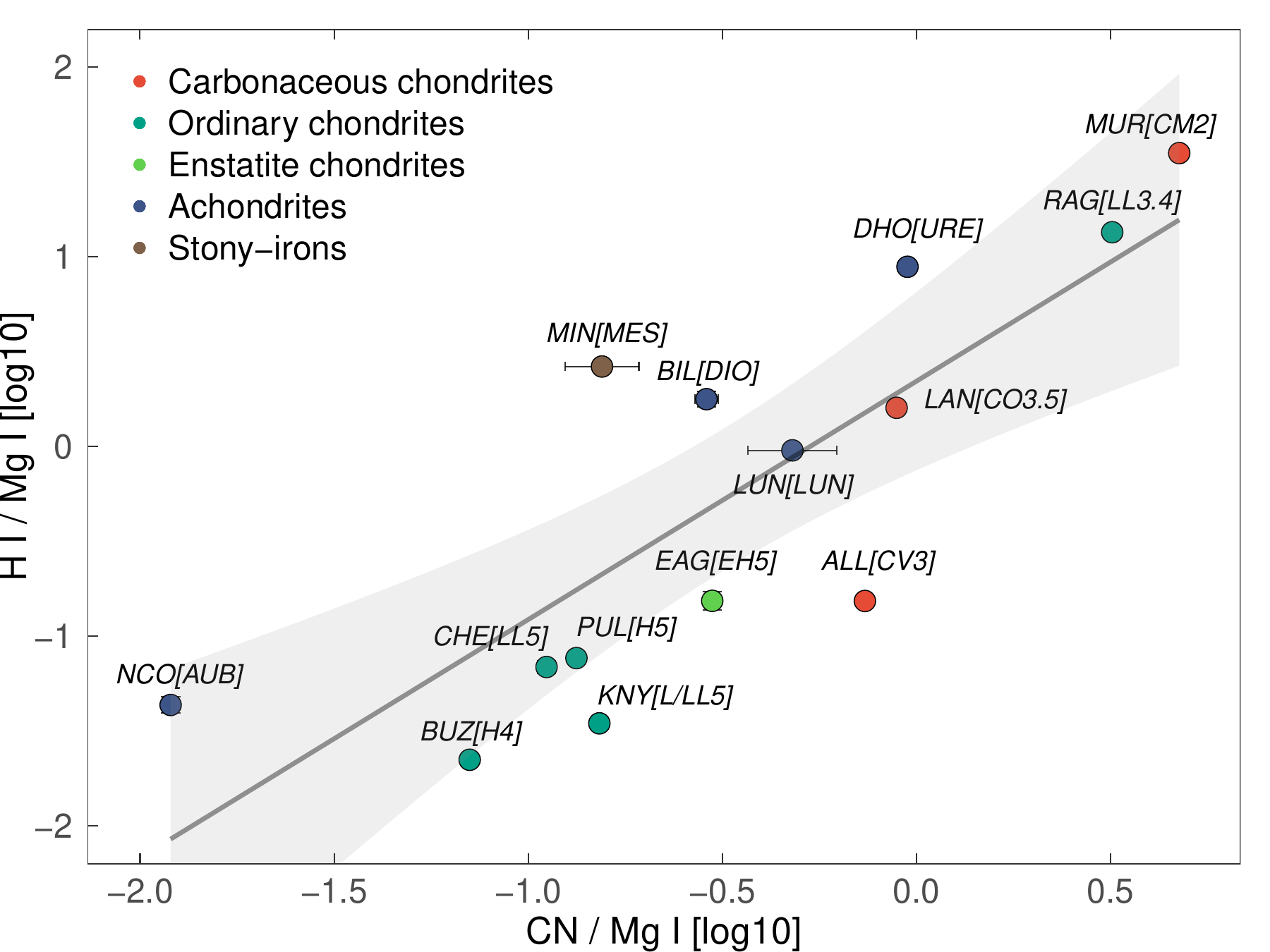}}
\caption{Observed H$\alpha$/Fe I-15 (upper plot) and H$\alpha$/Mg I-2 (lower plot) intensity ratio as a function of CN/Fe I-15 and CN/Mg I-2 intensity ratios in the emission spectra of ablated meteorites. The linear fit to the running average with standard error of the mean is shown by the gray line and area. Meteorite codes and measured relative intensity ratios can be found in Table \ref{table1}. The ordinary chondrites Pultusk, Buzzard Coulee, Chelyabinsk, and Knyahinya are plotted for context but their H$\alpha$ and CN intensities are too low for reliable confirmation. Meteorites in which no CN was detected and H$\alpha$ emission was dominantly caused by contamination from an external source are not plotted.} 
\label{HCN_FE}
\end{figure}

\begin{table}[]
\scriptsize
\centering
\caption{Measured CN/Fe I-15, CN/Mg I-2 and 388.3 nm/386.0 nm peak intensity ratios in Echelle spectra from simulated ablation of different meteorite classes. Each meteorite is designated with a name, corresponding class and abbreviation used in the presented plots. The given meteorites are in the order of the strongest CN emission. Meteorites with no CN are marked as NA. The 388.3 nm/386.0 nm peak intensity ratio was measured in all meteorites, although no CN signal was detected in some meteorites. In these cases, the signal at 388.3 nm originates from a very faint Fe I line (see discussion in Section \ref{CN/FeI-4_intensity_ratio}).} 
\begin{tabular}{@{}lccccc@{}}
\toprule
Meteorite      & Class        & Abbreviation        & CN/Fe I-15  & CN/Mg I-2   & \begin{tabular}{@{}c@{}}388.3 nm/386.0 nm \\ peak height \end{tabular} \\ \midrule
Murchison      & CM2          & MUR{[}CM2{]}        & 0.89 ± 0.02 & 4.75 ± 0.03 & 1.07                      \\
Ragland        & LL3.4        & RAG{[}LL3.4{]}      & 0.65 ± 0.03 & 3.19 ± 0.12 & 0.34                      \\
Dhofar 1575    & Ureilite     & DHO{[}URE{]}        & 0.58 ± 0.02 & 0.95 ± 0.05 & 0.81                      \\
Lancé          & CO3.5        & LAN{[}CO3.5{]}      & 0.24 ± 0.01 & 0.89 ± 0.01 & 0.18                      \\
Allende        & CV3          & ALL{[}CV3{]}        & 0.20 ± 0.01 & 0.74 ± 0.01 & 0.20                      \\
NWA 11303      & Lunar        & LUN{[}LUN{]}        & 0.20 ± 0.05 & 0.48 ± 0.13 & 0.16                      \\
Bilanga        & Diogenite    & BIL{[}DIO{]}        & 0.12 ± 0.01 & 0.29 ± 0.02 & 0.18                      \\
Norton County  & Aubrite      & NCO{[}AUB{]}        & 0.08 ± 0.01 & 0.01 ± 0.01 & 0.17                      \\
Eagle          & EH5          & EAG{[}EH5{]}        & 0.07 ± 0.01 & 0.30 ± 0.01 & 0.06                      \\
Pultusk        & H5           & PUL{[}H5{]}         & 0.04 ± 0.01 & 0.13 ± 0.01 & 0.13                      \\
Knyahinya      & L/LL5        & KNY{[}L/LL5{]}      & NA          & NA          & 0.04                      \\
Mincy          & Mesosiderite & MIN{[}MES{]}        & NA          & NA          & 0.06                      \\
Chelyabinsk    & LL5          & CHE{[}LL5{]}        & NA          & NA          & 0.04                      \\
Buzzard Coulee & H4           & BUZ{[}H4{]}         & NA          & NA          & 0.03                      \\
Sariçiçek      & Howardite    & SAR{[}HOW{]}       & NA          & NA          & 0.07                      \\
Tissint        & Shergottite  & TIS{[}SHE{]}       & NA          & NA          & 0.06                      \\
Kheneg Ljouâd  & LL5/6        & KLJ{[}LL5/6{]}     & NA          & NA          & 0.06                      \\
Stannern       & Eucrite      & STA{[}EUC{]}       & NA          & NA          & 0.05                      \\
Mount Joy      & Iron/IIAB    & MJO{[}IRON/IIAB{]} & NA          & NA          & 0.04                      \\
NWA 869        & L3-6         & NWA{[}L3-6{]}      & NA          & NA          & 0.03                      \\
Košice         & H5           & KOS{[}H5{]}        & NA          & NA          & 0.03                      \\
Mocs           & L5-6         & MOC{[}L5-6{]}      & NA          & NA          & 0.02                      \\ \bottomrule
\label{table1}
\end{tabular}
\end{table}

We have studied high-resolution Echelle spectra of 22 different meteorites obtained during their simulated ablation in plasma wind tunnel facility. The primary focus of this section is on the study of the occurrence and relative intensity of the CN band measured relative to the main meteor emission multiplets of Fe I-15 and Mg I-2 representing silicate and metallic components in meteoroids. These multiplets were selected as they are among the most universally observed features in visible-range meteor spectra. Additionally, we have studied the correlation between CN and H$\alpha$ near 656.3 nm since both features originate from volatiles embedded in meteorites and are potentially the best candidates for tracing water molecules and organic compounds in small solar system bodies \citep{10.1093/mnras/stac927}.

The generated free plasma flow at the beginning of each experiment enabled us to observe the plasma spectrum before moving the meteorite sample to the plasma flow, i.e. before the meteorite ablation started, allowing us to identify plasma lines and possible contribution to H emission from outside source. All spectra were thoroughly examined for possible H contamination, and four meteorite spectra with CN emission (Bilanga, Eagle, Lancé, and NWA 11303 meteorites) were confirmed with additional source of H emission. In the case of the Bilanga, Eagle and Lancé meteorite spectra, H emission was already observed before the meteorite insertion and increased after ablation started, indicating that some fraction of water molecules and organic compounds originate in these meteorites. The additional source of H was also detected in the spectrum of the NWA 11303 meteorite but without significant intensity change after meteorite ablation starts, pointing out the absence of the original H source from the meteorite. The external H source probably originated in the evaporated water of the internal cooling system of the plasma wind tunnel facility. No contamination of the detected H emission was found in the majority of the presented meteorite spectra. In addition, out of the meteorites used in the analysis, four finds (Dhofar 1575, Mincy, NWA 13303, and Ragland meteorite) may be to some degree affected by terrestrial weathering, whereas all the other tested meteorites are falls and, thus, much more pristine. The effects of the terrestrial weathering will be further examined based on the time evolution of meteorite spectra from individual frames.

In a previous work by \citet{10.1093/mnras/stac927}, based on a limited number of samples, a correlation was found between the intensity of the H$\alpha$ line and the CN band, which we here confirm based on an extended set of samples (namely Eagle [EH5], Bilanga [DIO], Lancé [CO3.5], Mincy [Mesosiderite], and Northwest Africa (NWA) 11303 [LUN]). Fig. \ref{HCN_FE} shows that meteorites with increased CN content also exhibit higher volatile H content. The strongest H and CN emissions were detected in the CM2 carbonaceous chondrite Murchison, which is, in fact, rich in hydrocarbons, amino-acids and water content $\sim$ 10 wt.\% \citep{1970Natur.228..923K, 1997Natur.389..265E, 2018GeCoA.239...17B}. We have also found stronger CN and H emissions in other carbonaceous chondrite meteorites, namely the CO3.5 Lancé and the CV3 Allende meteorites. The Allende meteorite contains on average $<$ 1 wt.\% water content \citep{2018GeCoA.239...17B}, which was manifested by a significantly lower H$\alpha$ line intensity compared to Murchison. Moreover, among all three tested carbonaceous chondrites, the Murchison meteorite (CM2) has the highest carbon content of 2.7 wt.\% (mean elemental abundance), followed by Lancé (CO3.5) and Allende (CV3) meteorites with 0.65 wt.\% and 0.27 wt.\% carbon content, respectively \citep{articlePearson}. Our results well reflect the real bulk elemental composition of these meteorites, as the Murchison meteorite exhibits the strongest CN/Fe I-15 ratio, followed by Lancé and Allende meteorites, as shown in Fig. \ref{HCN_FE} (upper panel). To account for the differences in the bulk composition of the individual meteorites, we display the intensities of H$\alpha$ line and CN relative to both Fe I and Mg I emission (Table \ref{table1}).

Within the group of achondrites, the strongest CN and H emissions were detected for the meteorite Dhofar 1575, belonging to the achondrite carbon-rich ureilite group. In ureilites, carbon is bound in the form of tiny grains of graphite and (nano)diamonds. Since this meteorite is a find, it is necessary to consider the potential influence of terrestrial weathering, although according to \citet{2014M&PS...49E...1R}, the weathering grade for this meteorite is low. The time evolution of the CN emission observed on monochromatic light curves (\ref{DHO_mono_curve_f43-80}) revealed continuous release of CN during the ablation, also supporting embedded source of CN within the meteorite sample.

Among other tested achondrites, relatively strong CN emission was detected for the lunar meteorite NWA 11303, mostly originating from the early stages of the meteorite ablation (see further discussion in Section \ref{light_curves}). On the contrary, the martian meteorite Tissint (Shergottite) did not exhibit any CN or H emission. Moderate CN and H emission was detected in the aubrite meteorite Norton County. Here we have found the most significant difference in the intensity of CN and H$\alpha$ relative to Fe I-15 and Mg I-2. The reason is its composition consisting of Mg-rich silicates and depleted in iron \citep{1985Metic..20..571E, 2011M&PS...46..284H}, which is reflected in low CN/Mg and H/Mg ratios and relatively high CN/Fe and H/Fe ratios, respectively (Fig. \ref{HCN_FE}). The Norton County aubrite contains $\sim$ 0.3 wt.\% water content \citep{2020Sci...369.1110P}. 

Moderate CN emission was also detected in the diogenite meteorite Bilanga. Interestingly, out of all the tested HED (howardite-eucrite-diogenite) meteorites, Bilanga is the only meteorite with detected CN as no CN was found in the eucrite Stannern or the howardite Sariçiçek. However, measurements of carbon isotopes which differentiate the presence of C caused by terrestrial contamination from indigenous content, confirmed the presence of indigenous carbon content in HED meteorites, including in the howardite Sariçiçek \citep{1997M&PS...32..863G, 2019M&PS...54.1495Y}. This level of carbon content however did not produce a detectable CN emission during the simulated ablation of the eucrite Stannern or the howardite Sariçiçek. To our knowledge, the carbon content of the Bilanga meteorite was not measured by previous authors, thus, we cannot compare with the other tested HED meteorites.

The H and CN line intensities are below the detection limit (log\textsubscript{10}(H/Fe I) $<$ -1.6 and log\textsubscript{10}(CN/Fe I) $<$ -1.3, respectively) in most of the tested ordinary chondrites, including Košice (H5), Pultusk (H5), Buzzard Coulee (H4), Mocs (L5-6), NWA 869 (L3-6), Knyahinya (L/LL5), Chelyabinsk (LL5), and Kheneg Ljouâd (LL5/6). Therefore, CN content was considered absent or unreliable in these tested meteorites.

While CN and H emissions were absent in most of the tested ordinary chondrites, they were surprisingly clearly detected in the LL3.4 ordinary chondrite Ragland. It has been reported that the terrestrial weathering altered metallic Fe, Ni and troilite to iron oxides and hydroxides \citep{1986Metic..21..217R} in the Ragland meteorite, and therefore we can assume a slight modification of its composition. However, Ragland has unusual mineralogical and chemical composition features for an LL ordinary chondrite. It has relatively high water content of 2.45 wt.\% and it is the least metamorphosed ordinary chondrite investigated in this study \citep{1986Metic..21..217R}. Therefore, the observed spectral features may also represents its original, atypical composition \citep{10.1093/mnras/stac927}.

We have found very faint CN band peak in the EH5 enstatite chondrite Eagle, which also corresponds with detected faint H$\alpha$ emission. Most of the detected CN emission originated from early stages of the meteorite ablation, implying source from the outer layers of the meteorite. Influence of the terrestrial weathering of the sample therefore cannot be excluded, although the sample originates from an observed fall. The water and carbon content of the Eagle meteorite are $\sim$ 0.5 wt.\% and $\sim$ 0.3 wt.\%, respectively \citep{1988Metic..23..379O, 1990Metic..25..323J}. It is believed that besides carbonaceous chondrites formed in the outer solar system as the main source of hydrated minerals delivered to Earth, enstatite chondrites from the inner solar system also contributed to the origin of Earth´s water \citep{2000M&PS...35.1309M, MARTY201256, 2020Sci...369.1110P}. Moreover, enstatite chondrites are considered to be the material from which the proto-Earth was formed, as they have identical isotopic abundances to terrestrial rocks \citep{JAVOY2010259, 2017Natur.541..521D}.

No CN emission was detected from the mesosiderite Mincy or the iron meteorite Mount Joy. Interestingly, we observed an onset of H emission in the early stages of the Mincy meteorite ablation with a gradual decrease and disappearance of the H$\alpha$ line. Since significant hydration is not assumed to be present in mesosiderites, the detection of H at the beginning of the ablation may reflect an effect of terrestrial weathering on the outer layers of the sample. We note that Mincy is a meteorite find.

\subsection{CN/Fe I-4 intensity ratio measurements and detection in lower resolution spectra}
\label{CN/FeI-4_intensity_ratio}

\begin{figure}
\centerline{\includegraphics[width=.95\columnwidth,angle=0]{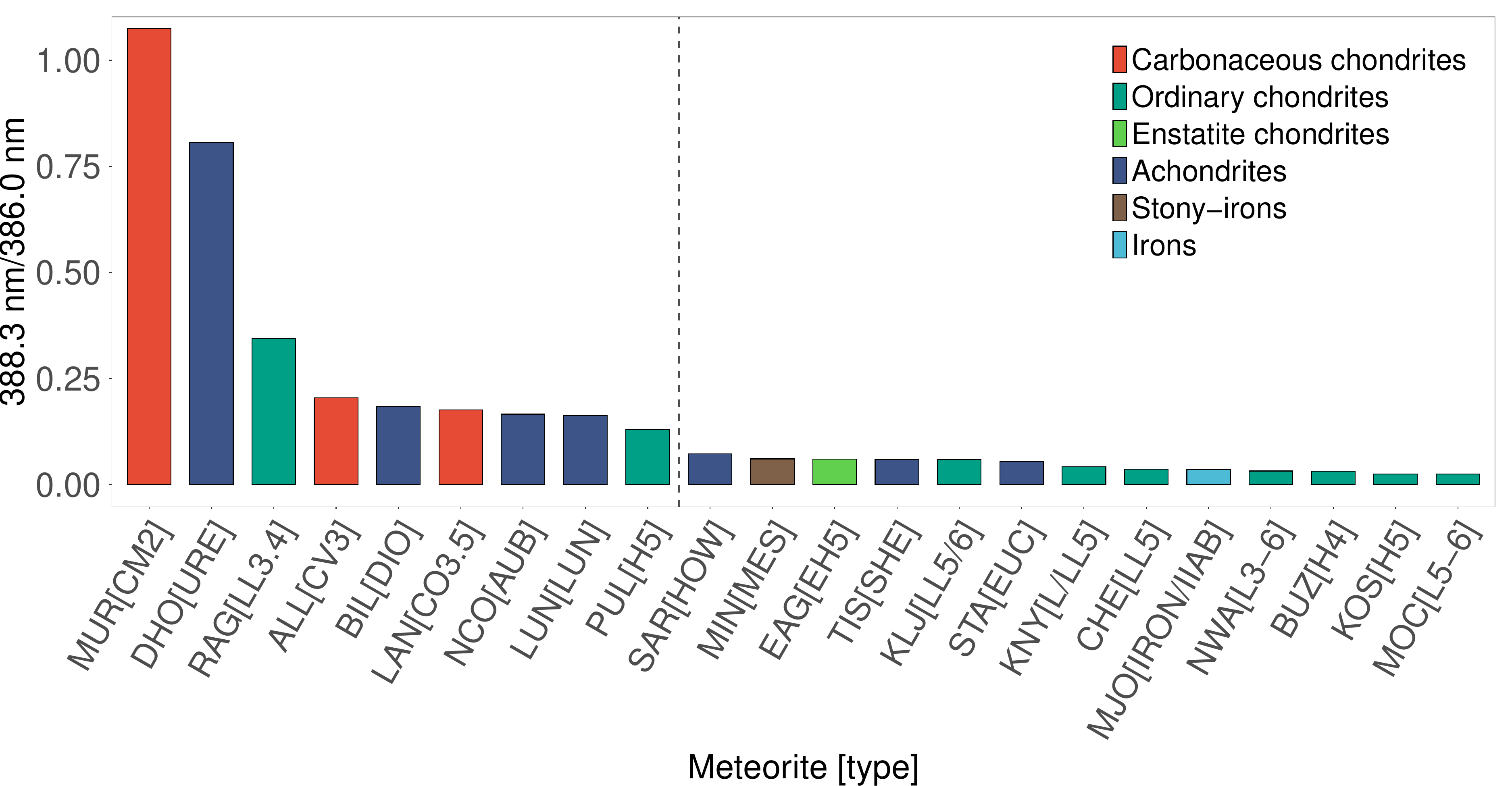}}
\caption{Bar chart of measured CN (388.3 nm)/Fe I (386.0 nm) intensity ratio from high-resolution Echelle spectra of all meteorites tested in the plasma wind tunnel. The meteorite codes can be found in Table \ref{table1}. The proposed dashed black boundary distinguishes meteorites with and without detected CN emission.} 
\label{3883nm_386nm}
\end{figure}

We have found that one of the most straightforward methods to recognize the presence of the CN band, which is also applicable to the lower resolution data, is by measuring the relative intensity ratio of the CN peak at 388.3 nm to the one line from the Fe I-4 multiplet positioned near 386.0 nm, as shown in Fig. \ref{3883nm_386nm} and Table \ref{table1}. Meteorites without CN present exhibit only very faint Fe I peak at the 388.3 nm. At 386.0 nm, all tested meteorites show a strong Fe I - 4 line peak. Without a considerable contribution of CN emission, the ratio between the Fe I lines at 388.3 nm/386.0 nm should remain relatively constant for different meteorite types, given that the ablation behavior of the sample is steady. Fig. \ref{3883nm_386nm} shows the distinction of meteorites in which this intensity ratio was increased compared to meteorites with no detected CN emission. The boundary for recognition of CN emission in our data seems to be the value of 388.3 nm/386.0 nm $\approx$ 0.1. In general, we did not find CN in meteorites with the intensity ratio below this value. However, we note that the distinction of CN contribution based on the 388.3 nm/386.0 nm intensity ratio must be considered carefully. In this work, the presence of CN was confirmed by also taking into account the overall intensity of the CN band relative to other element lines and studying the time evolution of its emission. 

The majority of all meteorites tested in the wind tunnel with apparent detection of CN emission belong to the group of carbonaceous chondrites (Murchison, Allende, Lancé) and distinct achondrites (Dhofar 1575, Bilanga, Norton County, and NWA 11303 meteorites). The surprising detection of CN in the spectrum of the ordinary chondrite Pultusk is assumed to be due to a contaminating source, as we only observed CN emission at the beginning of the ablation (see Fig. \ref{PUL_mono_curve_f19-93_CN_FeI4} and the discussion in Section \ref{light_curves}). 

\begin{table}[]
\centering
\caption{Measured 388.3 nm/386.0 nm peak intensity ratio and 388.3 nm/386.0 nm FWHM (Full Width at Half Maximum) ratio in AMOS-Spec-HR spectra of ablated meteorites demonstrating the main distinctive features of the detection of CN band applicable to the lower resolution meteor data (see also Fig. \ref{profiles}). The modeled lines had a Gaussian profile. The presented meteorites are in the order of the strongest CN emission. The CN band was not reliably detected in the Košice meteorite.}
\begin{tabular}{@{}lccc@{}}
\toprule
Meteorite   & Class        & \begin{tabular}{@{}c@{}}388.3 nm/386.0 nm \\ peak height\end{tabular}   &  \begin{tabular}{@{}c@{}}388.3 nm/386.0 nm \\ FWHM ratio \end{tabular}               \\ \midrule
Murchison   & CM2          & 0.80      & 0.70                      \\
Dhofar 1575 & Ureilite     & 0.77      & 0.68                      \\
Lancé       & CO3.5        & 0.62      & 1.34                      \\
Eagle       & EH5          & 0.69      & 1.35                      \\
Košice      & H5           & 0.59      & 1.60                      \\ \bottomrule
\label{table2}
\end{tabular}
\end{table}

The measurement of the 388.3 nm/386.0 nm intensity ratio presents a suitable method of identifying the presence of CN emission in lower resolution data. To validate this method, we measured the intensity ratio in the meteorite ablation spectra captured by the AMOS-Spec-HR spectrograph routinely used for meteor observations. Our results show that the presence of the CN band is accompanied by increasing the 388.3/386.0 nm peak intensity ratio and decreasing line width ratio of these two peaks due to the contribution of the surrounding CN peaks near 387.1 nm and 385.0 nm (Table \ref{table2}, Fig. \ref{profiles}). However, in the case of meteors, it is necessary to study these distinguishing features carefully, as the line emission depends on the flow enthalpy, i.e., on the entry speed (line intensity ratios may differ at different temperatures).

\subsection{Monochromatic light curves}
\label{light_curves}

\begin{figure}
\centerline{\includegraphics[width=.95\columnwidth,angle=0]{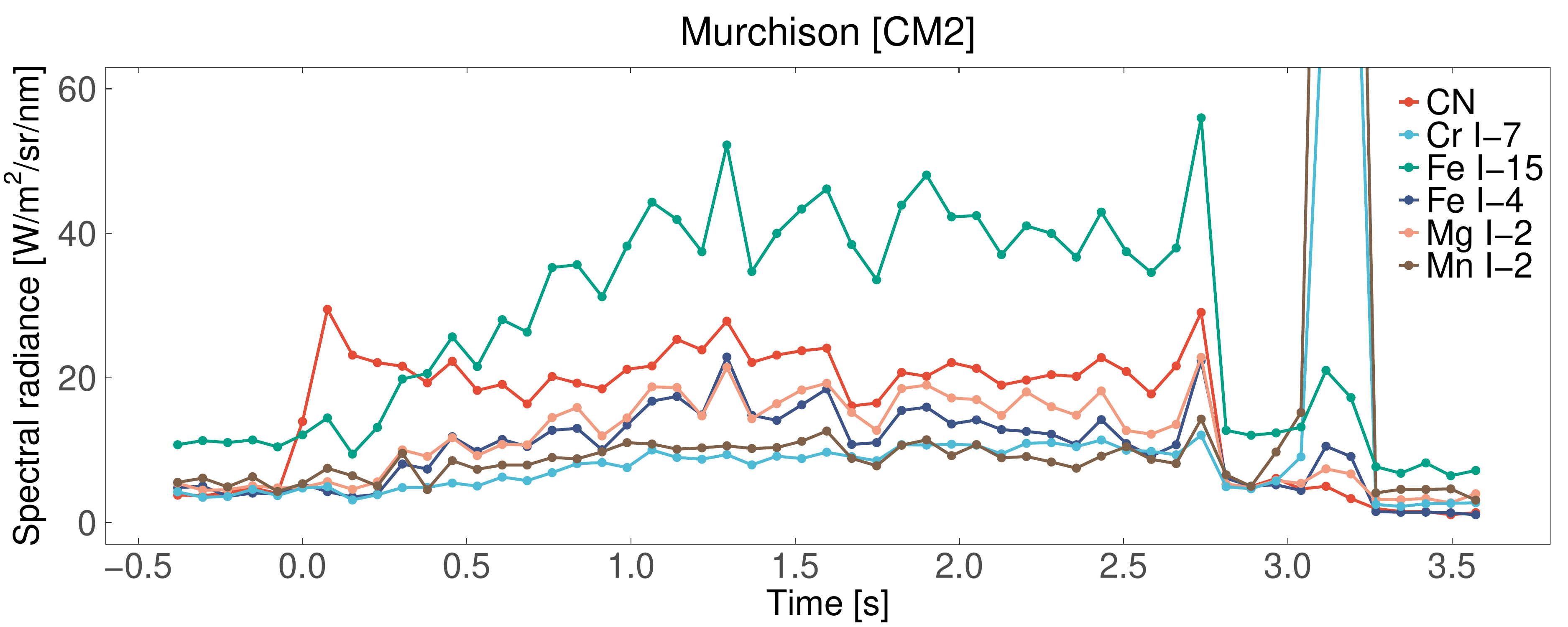}}
\caption{The monochromatic light curve of the CM2 carbonaceous chondrite meteorite Murchison demonstrating the early emission of CN and bright flare at the end of the ablation. Here, five frames before and after meteorite ablation are also plotted. The Fe I-4 monochromatic light curve is only represented by the intensity of the 388.6 nm line. The peaks of the Cr I -7 and Mn I-2 multiplets are outside the plot limits.} 
\label{MUR_mono_curve_f97-149_int60}
\end{figure}

Next, we studied the time evolution of the CN emission in the spectra of tested meteorites by analyzing their monochromatic light curves. We have found that the early sharp increase of CN intensity can be observed in the earliest stages of the meteorite ablation, along with the onset of the emission of Na, due to the volatility of its atoms and the low excitation potential of the Na lines. An example of the monochromatic light curve of the CM2 carbonaceous chondrite Murchison with the most notable CN emission is displayed in Fig. \ref{MUR_mono_curve_f97-149_int60}. Five frames before and after the meteorite ablation showing spectrum noise level are displayed for a better distinction of the onset of the CN emission. The value of 0.0 s on the x-axis corresponds to the start of the meteorite ablation, i.e. to the onset of Na lines which are the first to begin to radiate. The intensity of Fe I-4 multiplet is only represented by the intensity of one line at 388.6 nm, close to the strongest CN peak. One can note a different behavior in the time-dependent CN emission, which peaks in the early stages of ablation compared to other elements detected in the meteorite spectrum showing slow onset of emission. In the later stages of the meteorite ablation, the CN generally radiates in a similar trend as the other elements, including very short and subtle flares. Similar behavior of early strong radiation to that of CN emission was observed for the low excitation Na lines. Due to the saturation of Na lines in most frames of the Echelle spectra, the relative intensity of the Na I lines was not investigated in this work. An interesting behavior was observed in the monochromatic light curve of the Murchison meteorite as a bright flare after the final stages of the meteorite ablation with increased Cr I-7 and Mn I-2 intensity (near 107 $W/m^2/sr/nm$ and 170 $W/m^2/sr/nm$, respectively). This flare may be associated with a sudden release of a droplet of molten material with a specific composition consisting of chromium and manganese-bearing minerals.

Monochromatic light curves of other ablated meteorites plotted from the highest to the lowest CN intensity can be found in the \ref{appendix_A}, demonstrating different content of organic compounds and the time evolution of the CN emission. Slight variations are observed in the onset of CN emission among spectra of the different ablated meteorites. Meteorite spectra of Murchison, Dhofar 1575 and Eagle exhibit CN emission simultaneously with Na emission. In the spectra of Allende, Bilanga, Norton County, NWA 11303 and Ragland meteorites, the onset of CN emission was observed from the second frame (around 0.08 s of ablation) and from the third frame (about 0.17 s of ablation) in the case of Lancé meteorite. In addition, the light curve shape of CN emission for the Dhofar 1575 meteorite (\ref{DHO_mono_curve_f43-80}) is not characterized by a typically very steep initial increase of brightness but rather by a very slow, gradual increase and decrease in brightness, which does not indicate an effect of terrestrial weathering, but may rather reflect a heterogeneous distribution of carbon within the ureilite meteorite. However, a steep increase in CN intensity in the early stages of ablation and a subsequent steep decrease towards the noise level of the recording in the case of the lunar NWA 11303 meteorite (\ref{LUN_mono_curve_f22-57}) points to a source from the surface layer of the meteorite, which may also result from terrestrial weathering. We note that the weathering grade of this meteorite is low \citep{2020M&PS...55..460G}. In this case, the origin of the detected CN emission was not clearly resolved. Further laboratory analysis of the bulk composition of this meteorite could help resolve this issue.

\begin{figure}
     \centering 
     \begin{subfigure}[b]{0.8\textwidth}
         \centering
         \includegraphics[width=\textwidth]{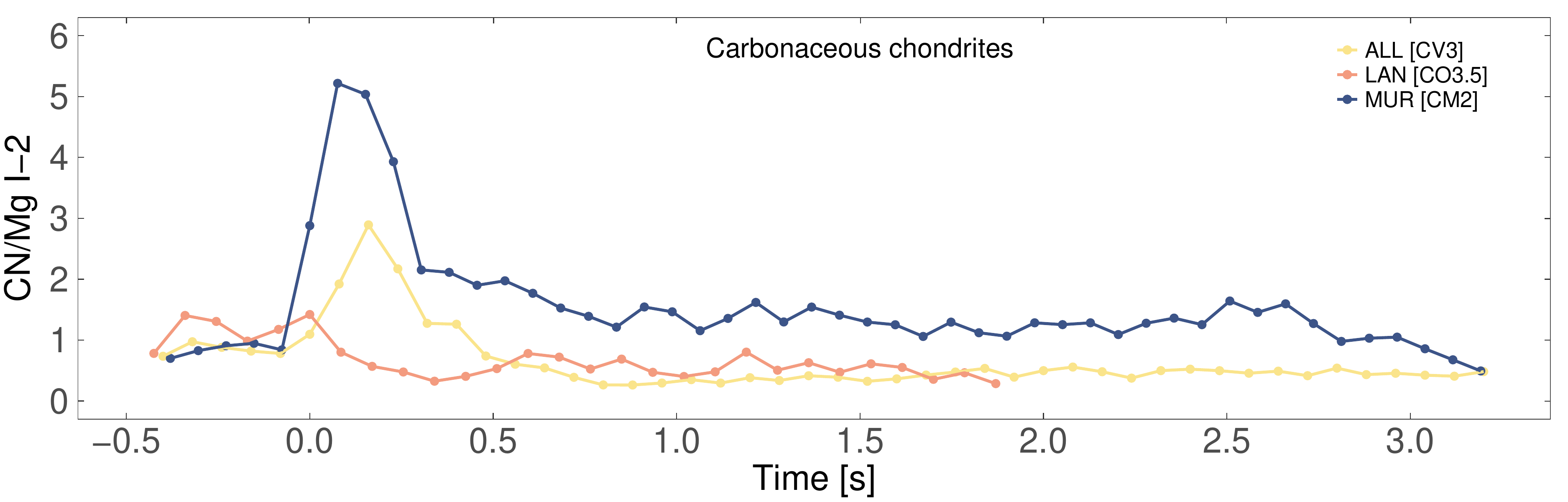}
         \caption{}
         \label{CNMgI2_vs_time_3,2s_carbonaceous_meteorites}
     \end{subfigure}
     \\
     \begin{subfigure}[b]{0.8\textwidth}
         \centering
         \includegraphics[width=\textwidth]{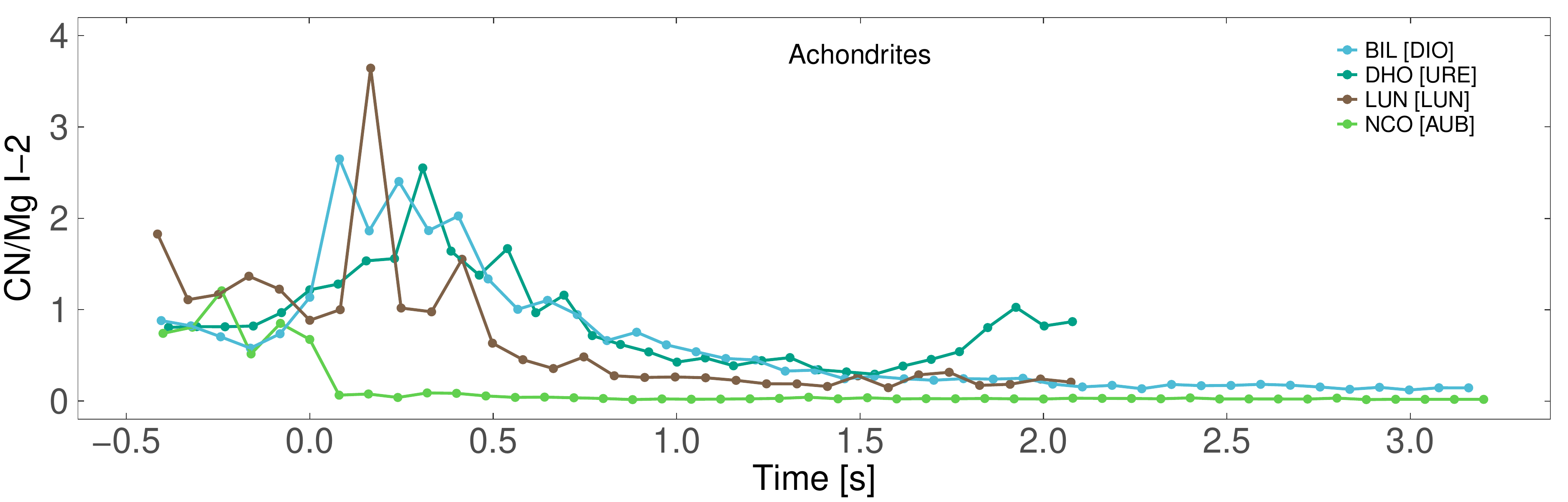}
         \caption{}
         \label{CNMgI2_vs_time_3,2s_achondrites}
     \end{subfigure}
     \\
     \begin{subfigure}[b]{0.8\textwidth}
         \centering
         \includegraphics[width=\textwidth]{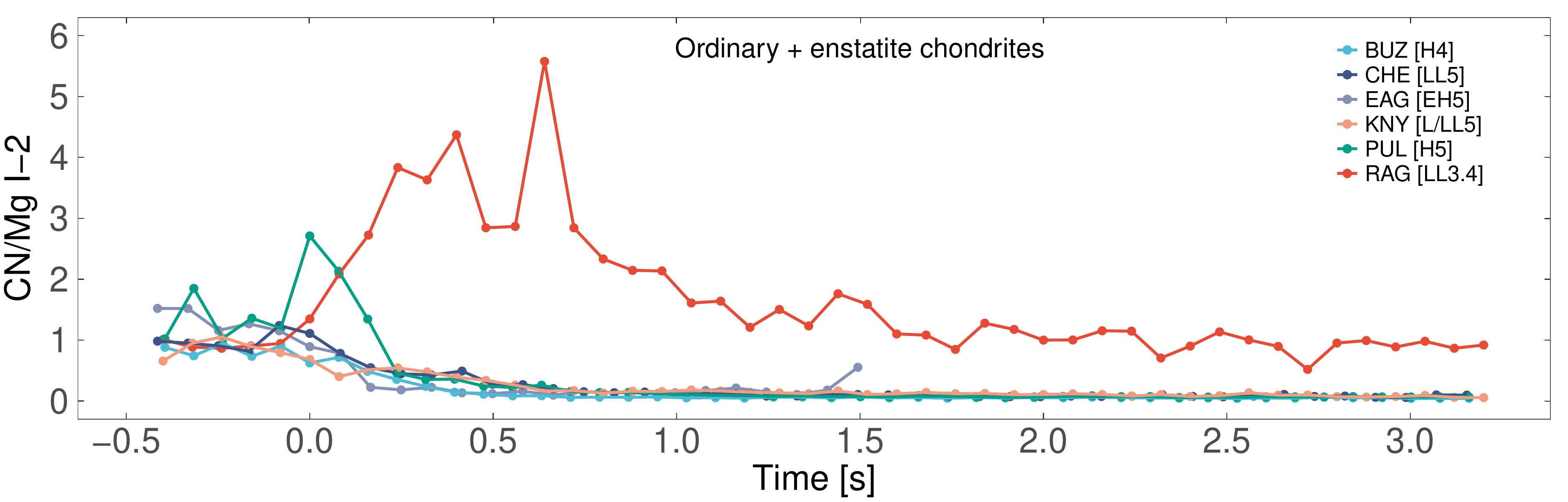}
         \caption{}
         \label{CNMg_vs_time_3,2s_ordinary_chondrites}
     \end{subfigure}
     \caption{Comparison of CN/Mg I-2 intensity ratios of meteorites as a function of time in the first 3.2 s of meteorite ablation for different compositional types. Five frames before meteorite ablation starts are also plotted. The CN intensity in the spectrum of Buzzard Coulee, Chelyabinsk, Eagle, and Knyahinya meteorites (c) was too low and likely represent faint Fe lines.}
     \label{CNMg_vs_time_3,2s}
\end{figure}

For further insight into the differential ablation of the studied meteorites, we measured the CN band peak intensity relative to the emission of other major atoms in individual frames. Fig. \ref{CNMg_vs_time_3,2s} displays the time-dependent CN/Mg I-2 intensity ratio for all meteorites with detected CN content. Similarly to Fig. \ref{MUR_mono_curve_f97-149_int60}, five frames are plotted before the meteorite ablation starts. Fig. \ref{CNMg_vs_time_3,2s} presents the evolution of CN emission relative to Mg I in the first 3.2 s of ablation for better visualization. However, as can be seen, the ablation duration of the Dhofar 1575, Eagle, Lancé, and NWA 11303 meteorite was shorter than for the other investigated meteorite samples.

The analyzed relative CN intensity ratios can be affected by the specific composition of the meteorite fragments, as observed by the high Mg content in the Norton County meteorite or the high Fe content in the Ragland meteorite (Fig. \ref{HCN_FE}). The CN line intensity measured relative to Fe I-15 in individual frames of meteorite ablation can be found in the \ref{appendix_A} for comparison. Regardless of the specific meteorite composition, a clear feature of the CN band is its strong peak in the early stages of the ablation and the subsequent sharp decrease in the CN/Mg and CN/Fe intensity ratio due to the gradual release of Mg and Fe dominating in the later stages. This result implies that, in lower-resolution spectra of real meteor observations, CN can be best detected in the early stages of the ablation in the upper atmosphere, before the strong emission of the surrounding Fe I begins. Sufficient instrument sensitivity and high frame rate may therefore play a key role for the detection of the early CN emission from meteors.

As already mentioned, the most notable CN emission was detected from carbonaceous chondrites. The light curve shapes of three ablated carbonaceous meteorites do not show significant variations with the exception of CN content (\ref{CNMgI2_vs_time_3,2s_carbonaceous_meteorites}). We observed the strongest CN peak in the CM2 Murchison meteorite, followed by the CV3 Allende and CO3.5 Lancé meteorites. In the case of the Lancé meteorite, it should be mentioned that the CN emission was observed starting from 0.17 s after the first detection of the meteorite spectrum (from the third frame) and the slight decrease of CN/Mg I-2 up to 0.5 s of ablation is caused by the significantly stronger Mg line (see also the monochromatic light curve of Lancé in \ref{LAN_mono_curve_f21-53}).

The time evolution of CN emission varies slightly between the achondrite meteorites (\ref{CNMgI2_vs_time_3,2s_achondrites}). The strongest peak of the CN band in the diogenite Bilanga was observed at around 0.08 s of the ablation process while in the lunar NWA 11303 and ureilite Dhofar 1575 meteorites, slightly later at 0.16 s and 0.3 s, respectively. The measured CN/Mg I-2 intensity ratio in the aubrite Norton County is continuously low (from the onset of CN emission at 0.08 s), but this fact is slightly affected by the relatively Mg-rich composition of this meteorite. When analyzed relative to the Fe I emission, the aubrite Norton County exhibits relative CN intensities closer to some other achondrites with notable CN presence (Fig. \ref{CNFeI15_vs_time_3,2s_achondrites}). The apparent increased trend of CN/Mg I-2 intensity ratio in Dhofar 1575 meteorite in the last stages of the meteorite ablation is the result of a significant decrease of Mg lines and relatively steadily slow CN release at the same time (see also the monochromatic light curve of Dhofar 1575 in \ref{DHO_mono_curve_f43-80}).

Our results suggest that ordinary and enstatite chondrites do not exhibit CN emission or only faint unrecognizable contribution near the background noise (\ref{CNMg_vs_time_3,2s_ordinary_chondrites}). As discussed earlier, the ablation of LL3.4-type Ragland meteorite was accompanied by strong CN emission with the most atypical time-depended behavior. The monochromatic light curve of Ragland is characterized by a gradually increasing CN/Mg I-15 intensity ratio in the first few frames and a subsequent very short and strong flare for the longest period of time compared to other meteorites with CN emission (see also the monochromatic light curve of Ragland in \ref{RAG_mono_curve_f52-131_int175}). This behavior can potentially be related to terrestrial weathering of this meteorite. Although the CN emission was not reliably resolved from the summed spectral profile of the H5 ordinary chondrite Pultusk (Fig \ref{HCN_FE}), we detected CN emission from the first second of the ablation (Fig. \ref{PUL_mono_curve_f19-93_CN_FeI4}). The observed distinct light curve of Pultusk likely reflects the presence of a surface layer rich in CN content. In this specific case, it is difficult to advocate the terrestrial weathering as this meteorite is a fall that was recovered quickly after its landing on Earth.

\section{Conclusions}
\label{conclusions}

We present here the first in-depth analysis of CN emission from a wide range of laboratory tested meteorites serving as asteroidal meteor analogues. The simulated ablation conditions correspond to an atmospheric flight of a slow meteoroid ($\sim$12\,km s\textsuperscript{-1}) at an altitude of approximately 80 km. The observed variations of CN emission in various meteorite types demonstrate that CN can be used as a diagnostic spectral feature of carbonaceous and relatively carbon-rich meteoroids. The strongest CN emission was found in carbonaceous chondrites (CM2, CV3 and CO3.5) and a C-rich ureilite. Moderate CN emission was found for diogenite, aubrite and lunar meteorite samples. Low CN contribution in the early stages of the ablation was found in an enstatite chondrite.

In general, the CN band was either absent or not clearly detected in most of the ablated ordinary chondrites with the exception of the Ragland meteorite (LL3.4), consisting of moderately weathered material with originally atypical composition for an ordinary chondrite. CN was not detected in the tested eucrite, howardite, martian shergottite, mesosiderite and iron meteorites.

Our results point out strong correlation between CN and H emission and suggest that both volatile features are suitable to trace contents of organic matter and water molecules present in meteoroids. While this study only focus on analogues of asteroidal meteors, our previous survey \citep{10.1093/mnras/stac927} pointed out strong H emission as a marker of the high volatile contents in cometary meteoroids.

The analysis of monochromatic light curves of ablated meteorites has shown that CN emission can be best recognized in the early stages of the meteorite ablation, before the onset of surrounding Fe I lines. For application in lower resolution meteor observations, we therefore suggest that efficient detection of CN can be achieved during the early stages of meteor ablation in the upper atmosphere. Additionally, using lower resolution data from the meteor spectrograph AMOS we found that the measurement of the intensity ratio and line width ratio of the CN band peak near 388.3 nm to the Fe I-4 line peak near 386.0 nm can indicate the contribution of CN in lower resolution spectra dominated by surrounding iron lines.

Our results suggest that terrestrial weathering of meteorites can affect their spectral signature including potentially affecting the tracers of water molecules and organic content. Such effects, typically detected in the samples that were more weathered meteorite finds, were resolved in specific monochromatic light curves showing CN or H emission only in early stages of the ablation. This may be explained by a contaminant source of carbon on the surface layers of the meteorite. Nevertheless, the strong spectral distinction between the carbon-rich materials, specific achondrites and ordinary chondrites confirms that the studied diagnostic spectral features correlate with their bulk composition and thus can be used to trace the original contents of organic compounds in meteoroids.

\section*{Acknowledgements}

We are thankful to the High Enthalpy Flow Diagnostics Group (HEFDiG) team of the Institute of Space Systems, University of Stuttgart for carrying out the meteorite ablation experiments. This work was supported by ESA grants under contracts No. 4000128930/19/NL/SC and No. 4000140012/22/NL/SC/rp, the Slovak Research and Development Agency grant APVV-16-0148, the Slovak Grant Agency for Science grant VEGA 1/0218/22, and the Comenius University Grant G-21-193-00 and G-22-145-00. J. Vaubaillon was supported by CNES, the French space agency, in the framework of the MALBEC project. We particularly thank all colleagues from HEFDiG in Stuttgart who supported and inspired the MetSpec campaigns. G. Batic (NHMW) is thanked for the preparation of the meteorite samples.

\section*{Data Availability}

The spectral data of presented ablated meteorites will be made available upon a reasonable request.

\appendix

\section{Additional monochromatic light curves}
\label{appendix_A}

\begin{figure}[H]
     \centering
     \begin{subfigure}[b]{0.9\textwidth}\ContinuedFloat
         \centering
         \includegraphics[width=\textwidth]{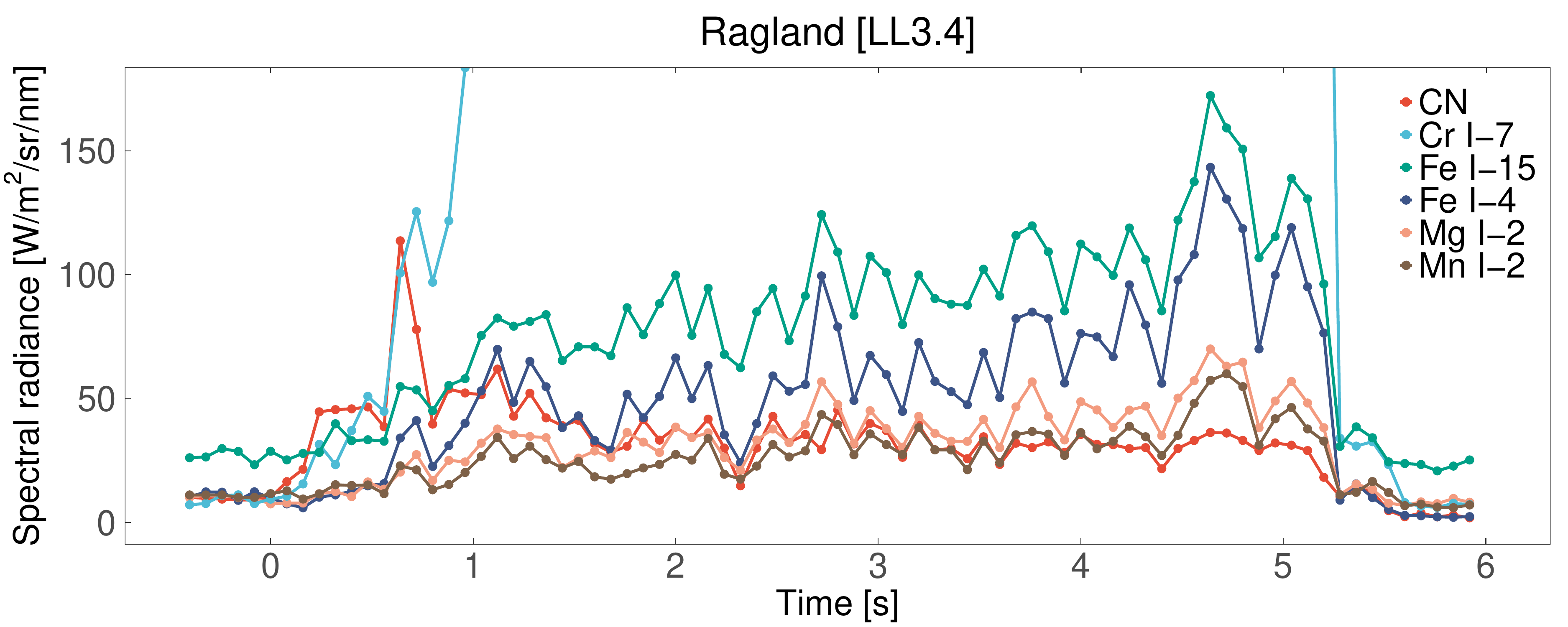}
         \caption{}
         \label{RAG_mono_curve_f52-131_int175}
     \end{subfigure}
     \\
     \begin{subfigure}[b]{0.9\textwidth}\ContinuedFloat
         \centering
         \includegraphics[width=\textwidth]{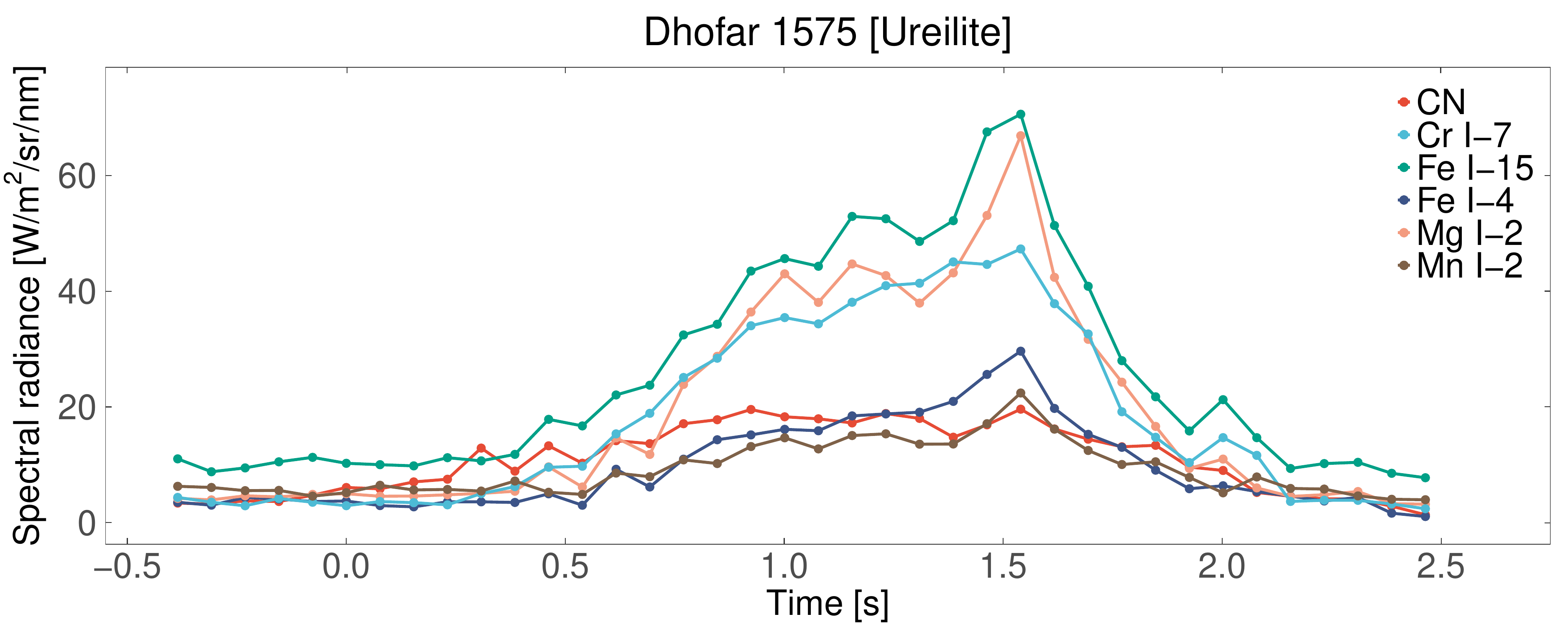}
         \caption{}
         \label{DHO_mono_curve_f43-80}
     \end{subfigure}
     \\
     \begin{subfigure}[b]{0.9\textwidth}\ContinuedFloat
         \centering
         \includegraphics[width=\textwidth]{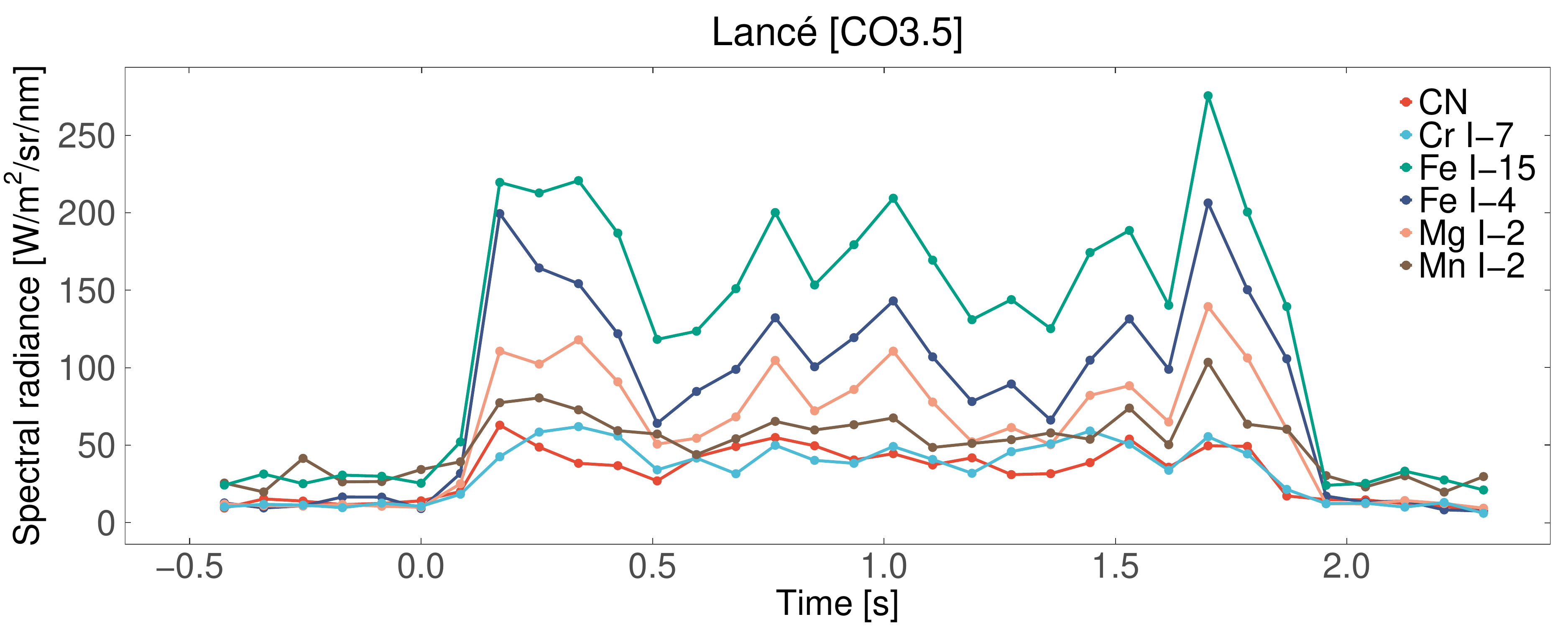}
         \caption{}
         \label{LAN_mono_curve_f21-53}
     \end{subfigure}
     \\
     \end{figure}
     
     \begin{figure}\ContinuedFloat
     \centering
    
     \begin{subfigure}[b]{0.9\textwidth}\ContinuedFloat
         \centering
         \includegraphics[width=\textwidth]{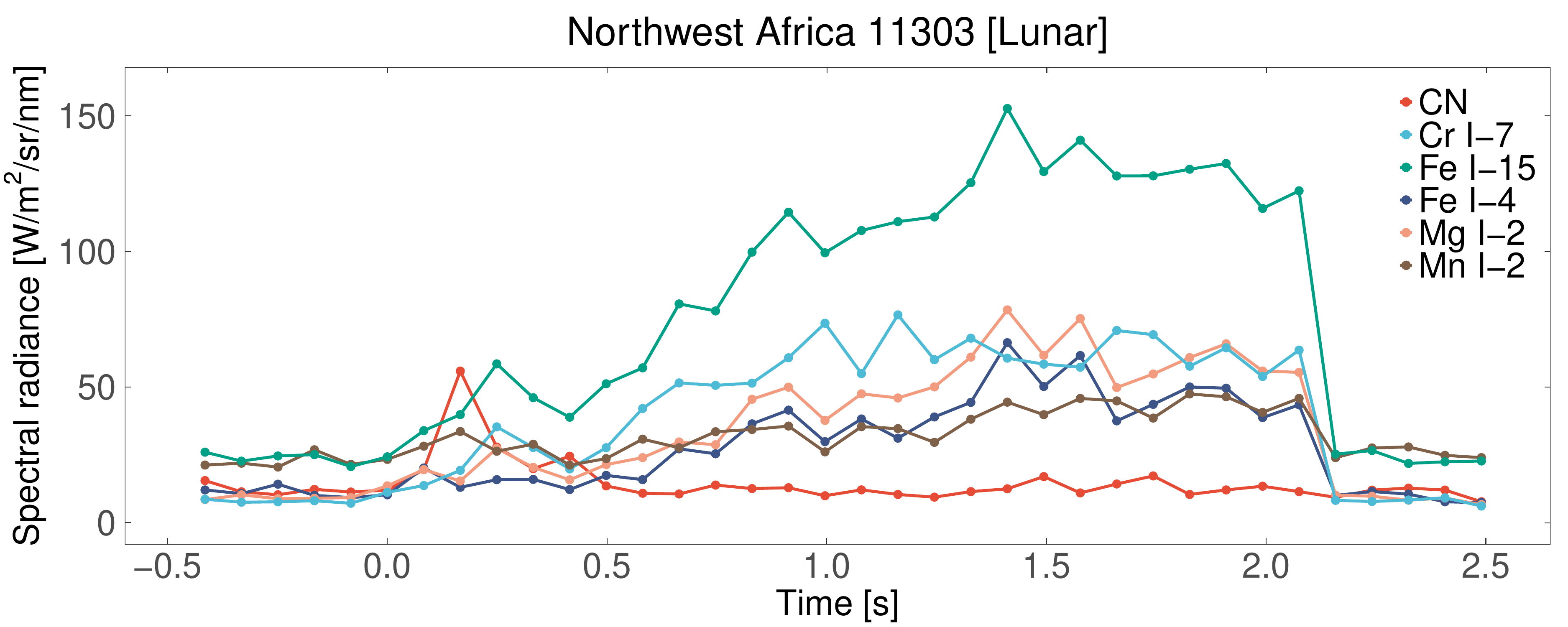}
         \caption{}
         \label{LUN_mono_curve_f22-57}
     \end{subfigure}
     \\
     \begin{subfigure}[b]{0.9\textwidth}\ContinuedFloat
         \centering
         \includegraphics[width=\textwidth]{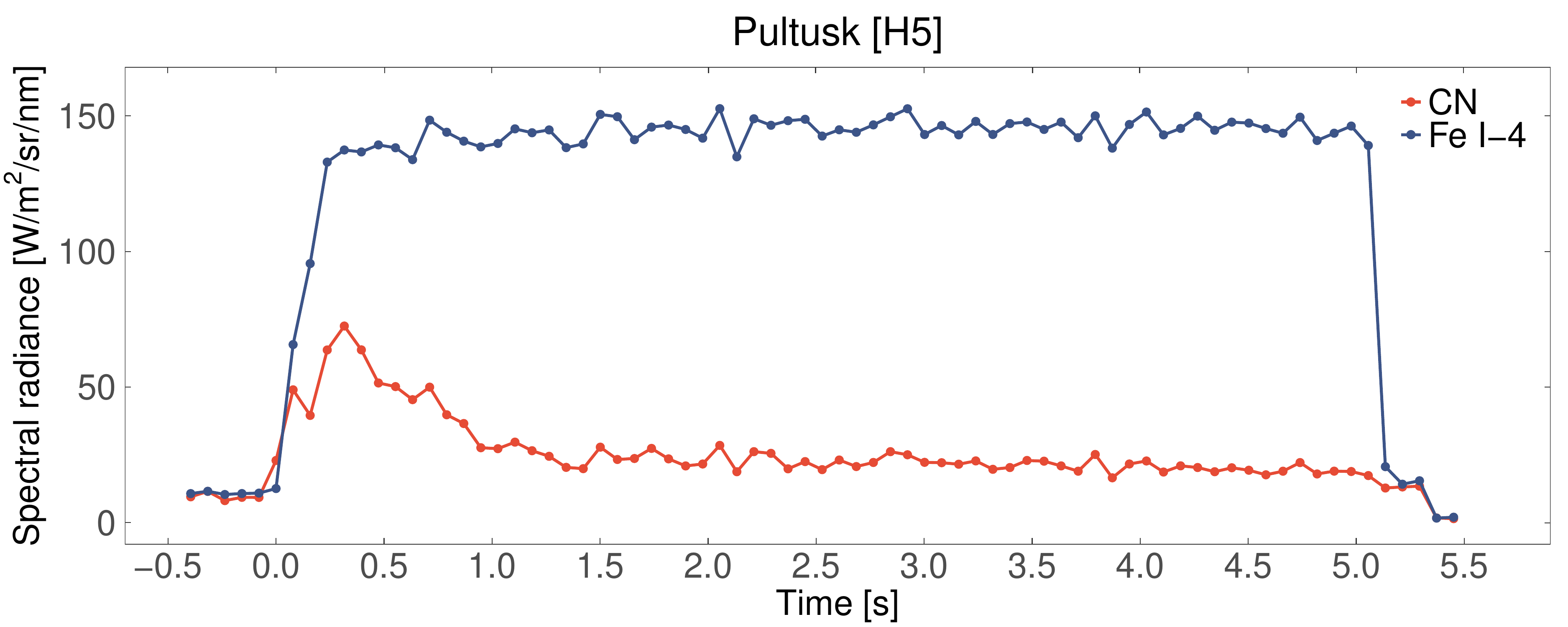}
         \caption{}
         \label{PUL_mono_curve_f19-93_CN_FeI4}
     \end{subfigure}
     \\
     \begin{subfigure}[b]{0.9\textwidth}\ContinuedFloat
         \centering
         \includegraphics[width=\textwidth]{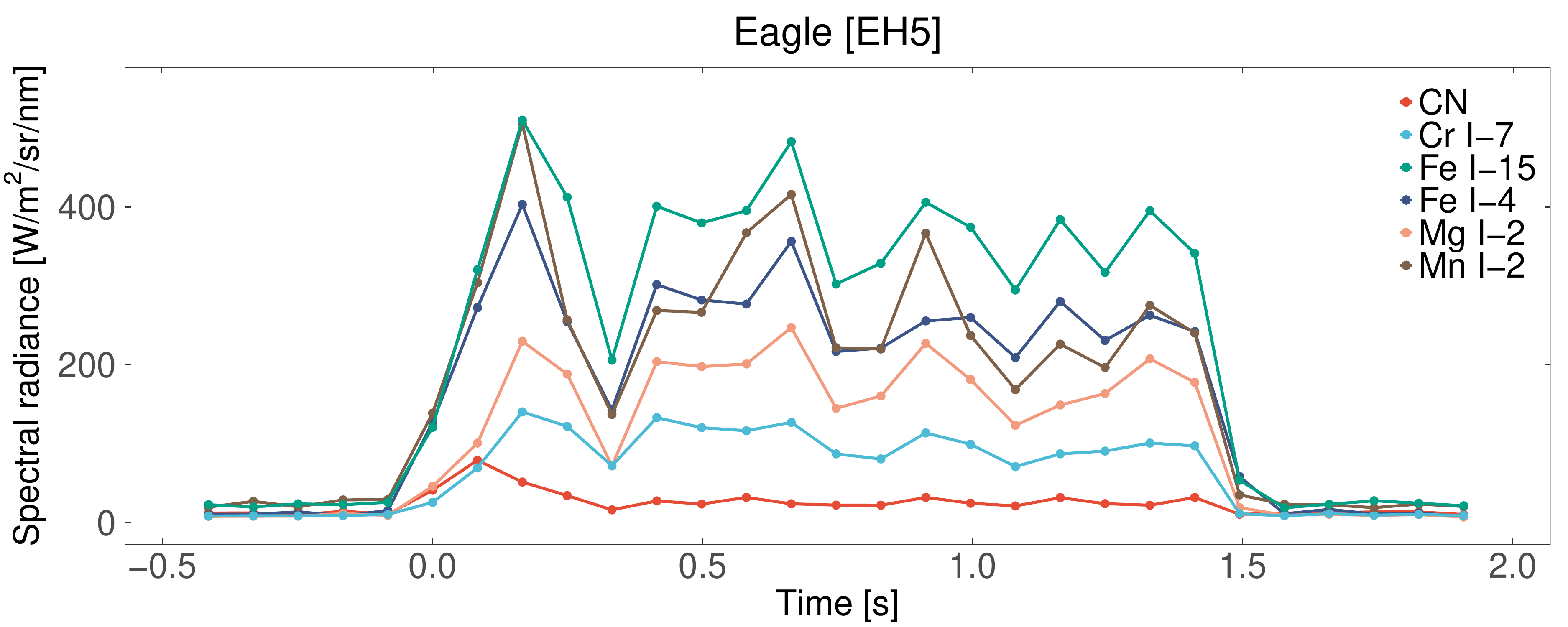}
         \caption{}
         \label{EAG_mono_curve_f54-82}
     \end{subfigure}
     \caption{Monochromatic light curves of six meteorites demonstrating different evolution of time-depended CN emission and content of organics in meteorites of different types in the order of the strongest CN intensity: Ragland [LL3.4], Dhofar 1575 [Ureilite], Lancé [CO3.5], Northwest Africa 11303 [Lunar], Pultusk [H5], and Eagle [EH5]. Five frames before and after meteorite ablation are also plotted. The intensity of Cr I-7 multiplet for Ragland is outside the plot. Only the monochromatic light curve of CN and Fe I-4 is plotted for Pultusk. Other lines are not plotted for better visibility.}
\end{figure}

\begin{figure}[H]
     \centering
     \begin{subfigure}[b]{0.8\textwidth}
         \centering
         \includegraphics[width=\textwidth]{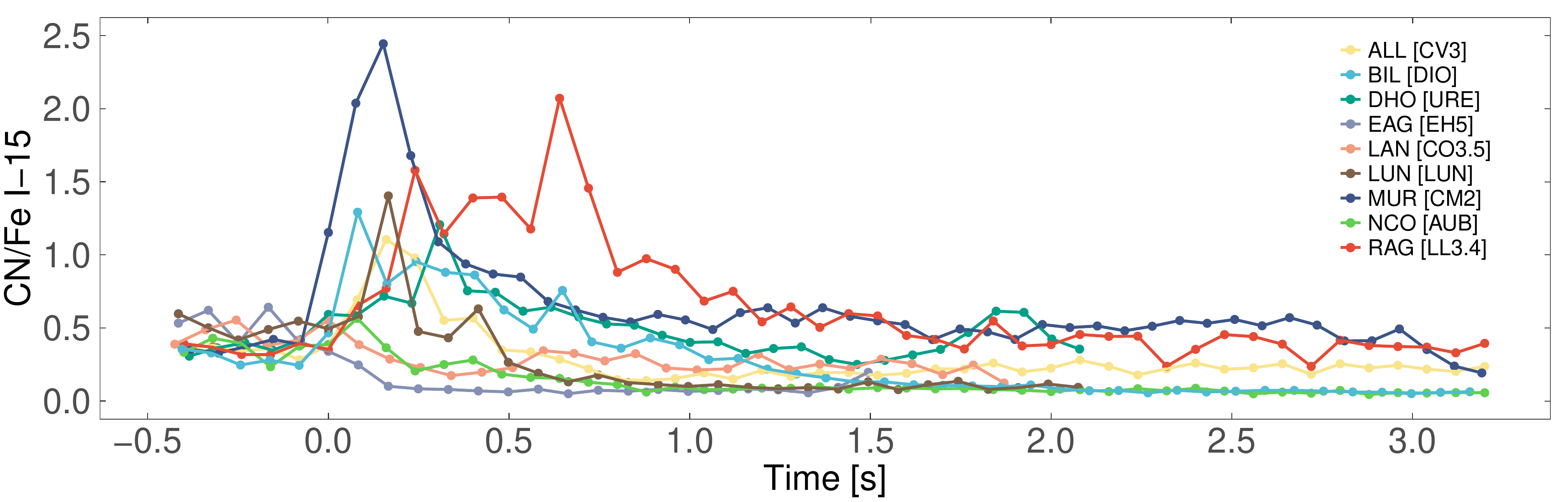}
         \caption{}
         \label{CNFeI15_vs_time_3,2s}
     \end{subfigure}
     \\
     \begin{subfigure}[b]{0.8\textwidth}
         \centering
         \includegraphics[width=\textwidth]{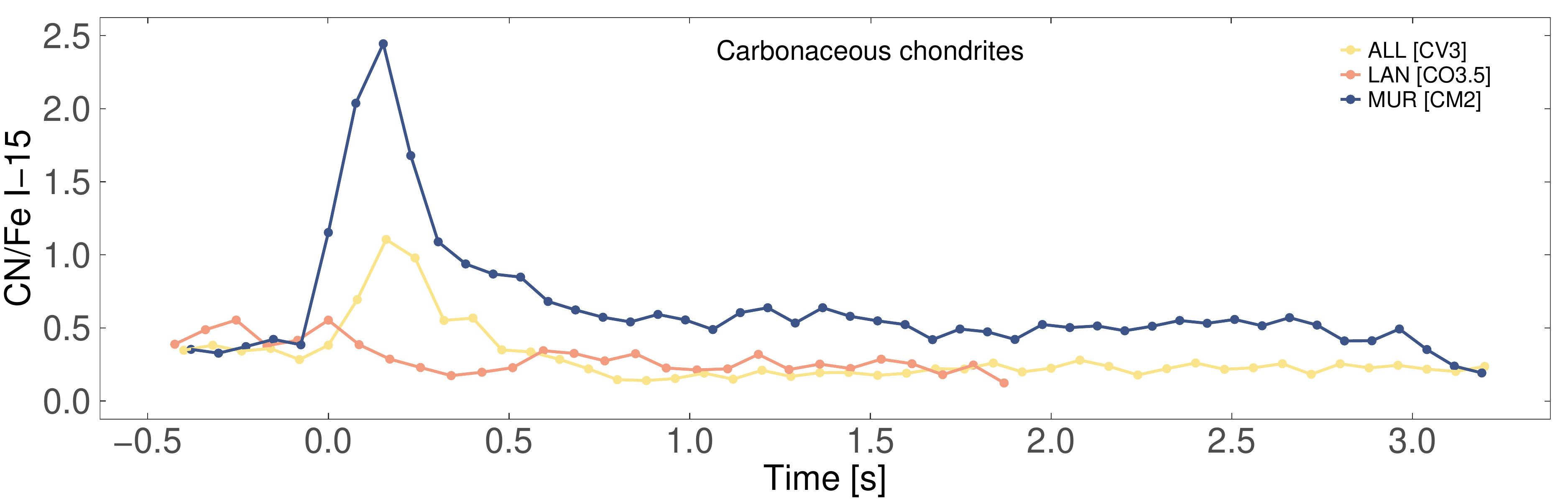}
         \caption{}
         \label{CNFeI15_vs_time_3,2s_carbonaceous_meteorites}
     \end{subfigure}
     \\
     \begin{subfigure}[b]{0.8\textwidth}
         \centering
         \includegraphics[width=\textwidth]{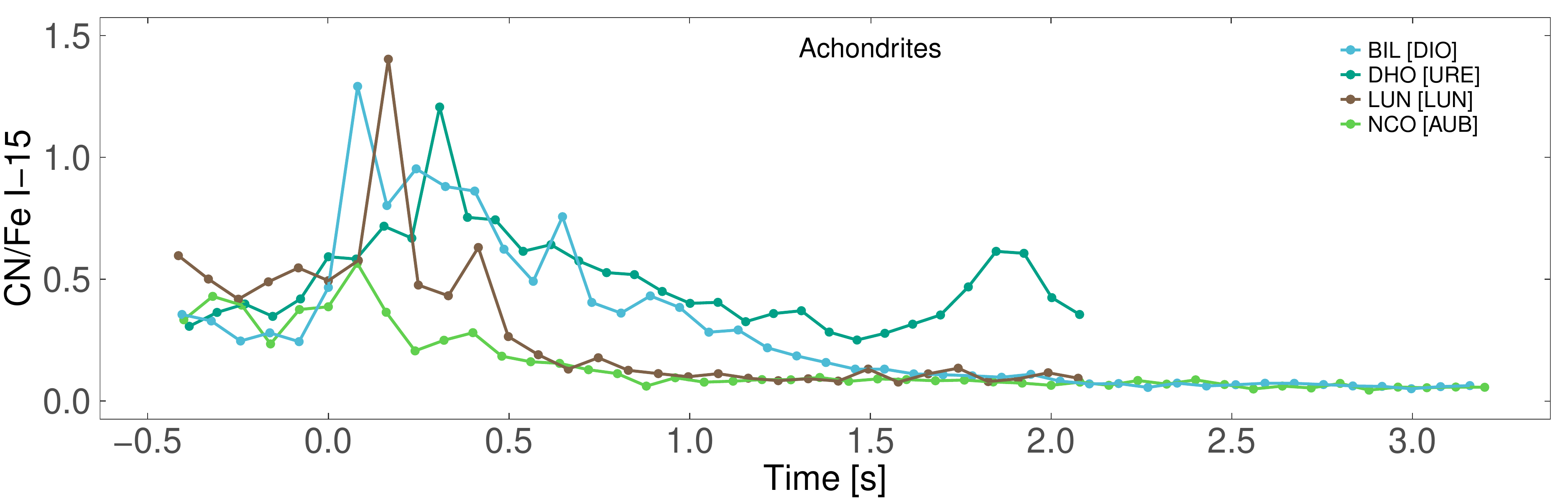}
         \caption{}
         \label{CNFeI15_vs_time_3,2s_achondrites}
     \end{subfigure}
     \\
     \begin{subfigure}[b]{0.8\textwidth}
         \centering
         \includegraphics[width=\textwidth]{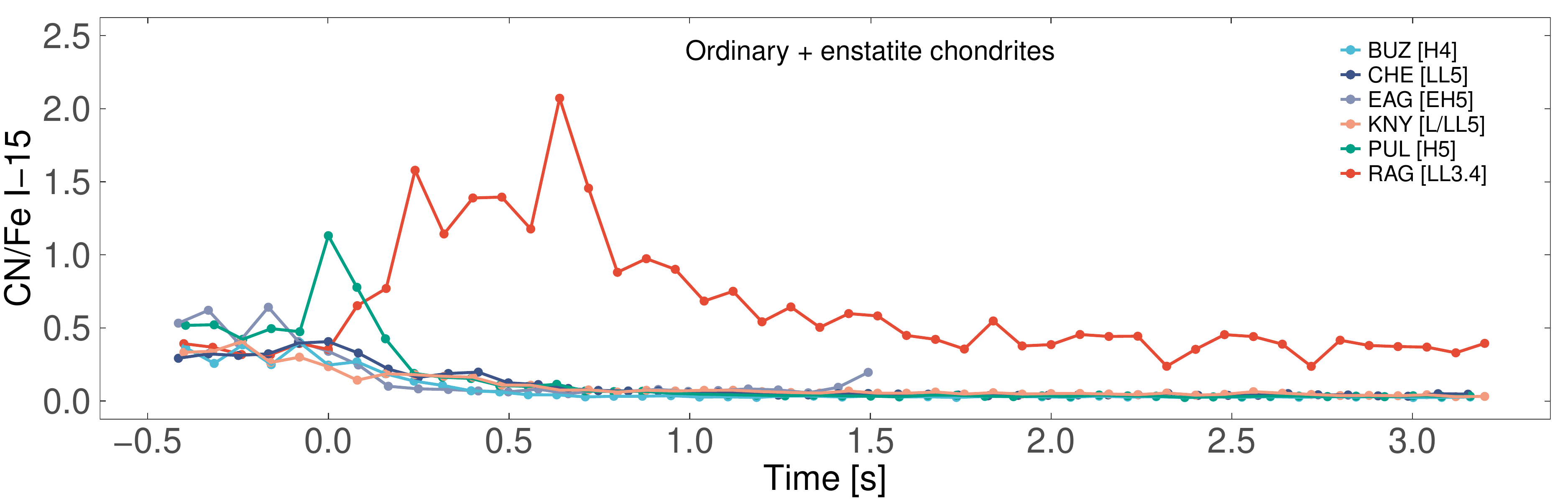}
         \caption{}
         \label{CNFeI15_vs_time_3,2s_ordinary_chondrites}
     \end{subfigure}
     \caption{The observed CN/Fe I-15 intensity ratio as a function of time in the first 3.2 s of meteorite ablation for all laboratory-tested meteorites with detected CN emission (a). Comparison of CN/Fe I-15 intensity ratios of meteorites with different compositional types is displayed in (b), (c) and (d). Five frames before meteorite ablation starts are also plotted. The CN intensity in the spectrum of Buzzard Coulee, Chelyabinsk, Eagle, and Knyahinya meteorites (d) was too low and may represent faint Fe lines.}
\end{figure}

\bibliographystyle{elsarticle-harv} 
\bibliography{references}





\end{document}